\documentclass[onecolumn,draftclsnofoot,nofonttune,11pt]{IEEEtran}
\usepackage{amsmath}
\usepackage{amssymb}
\usepackage{amsfonts}
\usepackage{amsthm}
\usepackage{mathtools}
\mathtoolsset{%
}

\usepackage[utf8]{inputenc}
\usepackage[T1]{fontenc}

\usepackage[varg]{txfonts}
\let\mathbb=\varmathbb

\usepackage{dsfont}

\usepackage[%
cal=cm,
]
{mathalfa}

\usepackage[dvipsnames,svgnames]{xcolor}
\colorlet{MyBlue}{DodgerBlue!40!Black}
\colorlet{MyGreen}{DarkGreen!85!Black}

\usepackage{subfigure}
\usepackage{tikz}
\usepackage{acronym}
\usepackage{balance}
\usepackage{latexsym}
\usepackage{paralist}
\usepackage{xspace}

\usepackage[numbers,sort&compress]{natbib}

\usepackage{hyperref}
\hypersetup{
draft=false,
colorlinks=true,
linktocpage=true,
pdfstartview=FitH,
breaklinks=true,
pdfpagemode=UseNone,
pageanchor=true,
pdfpagemode=UseOutlines,
plainpages=false,
bookmarksnumbered,
bookmarksopen=false,
bookmarksopenlevel=1,
hypertexnames=true,
pdfhighlight=/O,
urlcolor=MyBlue,linkcolor=MyBlue,citecolor=MyGreen, % <--- for screen
pdftitle={},
pdfauthor={},
pdfsubject={},
pdfkeywords={},
pdfcreator={pdfLaTeX},
pdfproducer={LaTeX with hyperref}
}

\newcommand{\bA}{\mathbf{A}}

\newcommand{\bD}{\mathbf{D}}

\newcommand{\bG}{\mathbf{G}}
\newcommand{\bH}{\mathbf{H}}
\newcommand{\bI}{\mathbf{I}}

\newcommand{\bQ}{\mathbf{Q}}

\newcommand{\bV}{\mathbf{V}}
\newcommand{\bW}{\mathbf{W}}
\newcommand{\bX}{\mathbf{X}}
\newcommand{\bY}{\mathbf{Y}}
\newcommand{\bZ}{\mathbf{Z}}

\newcommand{\bi}{\mathbf{i}}
\newcommand{\bj}{\mathbf{j}}
\newcommand{\bk}{\mathbf{k}}

\newcommand{\bx}{\mathbf{x}}

\newcommand{\CC}{\mathbb{C}}
\newcommand{\R}{\mathbb{R}}

\newcommand{\N}{\mathbb{N}}

\DeclareMathOperator{\bigoh}{\mathcal O}

\DeclareMathOperator{\diag}{diag}

\DeclareMathOperator{\ex}{\mathbb{E}}
\DeclareMathOperator{\grad}{\nabla}

\DeclareMathOperator{\one}{\mathds{1}}

\DeclareMathOperator{\prob}{\mathbb{P}}
\DeclareMathOperator{\rank}{rank}

\DeclareMathOperator{\tr}{tr}

\newcommand{\dd}{\:d}

\newcommand{\eps}{\varepsilon}

\newcommand{\mgeq}{\succcurlyeq}
\newcommand{\ml}{\prec}

\newcommand{\nhd}{U}

\newcommand{\wilde}{\widetilde}

\providecommand\given{} % provides an empty command for the delimiters below

\DeclarePairedDelimiter{\braces}{\{}{\}}
\DeclarePairedDelimiter{\bracks}{[}{]}
\DeclarePairedDelimiter{\parens}{(}{)}

\DeclarePairedDelimiter{\abs}{\lvert}{\rvert}
\DeclarePairedDelimiter{\norm}{\lVert}{\rVert}

\DeclarePairedDelimiterX{\braket}[2]{\langle}{\rangle}{#1\mathopen{}\delimsize\vert\mathopen{}#2}

\DeclarePairedDelimiterX{\product}[2]{\langle}{\rangle}{#1,#2}
\DeclarePairedDelimiterX{\setdef}[2]{\{}{\}}{#1:#2}

\DeclarePairedDelimiterXPP{\exclude}[1]{\mathopen{}\setminus}{\{}{\}}{}{#1}

\DeclarePairedDelimiterXPP{\probof}[1]{\prob}{(}{)}{}{%
\renewcommand\given{\nonscript\:\delimsize\vert\nonscript\:\mathopen{}}
#1}

\DeclarePairedDelimiterXPP{\exof}[1]{\ex}{[}{]}{}{%
\renewcommand\given{\nonscript\:\delimsize\vert\nonscript\:\mathopen{}}
#1}

\DeclarePairedDelimiterXPP{\trof}[1]{\tr}{[}{]}{}{#1}
\DeclarePairedDelimiterXPP{\logof}[1]{\log}{(}{)}{}{#1}

\newcommand{\dis}{\displaystyle}
\newcommand{\txs}{\textstyle}

\newcommand{\insum}{\sum\nolimits}

\newcommand{\as}{\textup(a.s.\textup)\xspace}

\usepackage[textwidth=30mm]{todonotes}
\usepackage{algorithm2e_extension}
\usepackage{algorithm}
\SetKwBlock{Repeat}{Repeat}{}
\newcommand{\kwd}[1]{\textsf{\bfseries#1}}

\newtheorem{theorem}{Theorem}
\newtheorem{corollary}{Corollary}
\newtheorem*{corollary*}{Corollary}

\newtheorem{proposition}{Proposition}

\theoremstyle{definition}
\newtheorem{definition}{Definition}
\newtheorem*{definition*}{Definition}

\theoremstyle{remark}
\newtheorem{remark}{Remark}
\newtheorem*{remark*}{Remark}

\newcommand{\feas}{\boldsymbol{\mathcal{X}}}

\newcommand{\dnorm}[1]{\norm{#1}_{\infty}}

\newcommand{\breg}{D_{\mathrm{KL}}}

\newcommand{\fench}{F}

\newcommand{\player}{i}
\newcommand{\players}{N}
\newcommand{\playerset}{\mathcal{\players}}
\newcommand{\playeralt}{j}
\newcommand{\play}{\player} 		% alias for player
\newcommand{\plays}{\players}		% alias for players
\newcommand{\playset}{\playerset}	% alias for player set
\newcommand{\playalt}{\playeralt}		% alias for alt player

\newcommand{\act}{\bX}

\newcommand{\eq}{\act^{\ast}}

\newcommand{\pay}{u}
\newcommand{\payv}{\bV}
\newcommand{\payveq}{\payv^{\ast}}
\newcommand{\vbound}{V}

\newcommand{\game}{\mathcal{G}}

\newcommand{\dkl}{D_{\textup{KL}}}

\newcommand{\step}{\gamma}

\newcommand{\carrier}{s}
\newcommand{\carriers}{S}

\newcommand{\effH}{\wilde\bH}
\newcommand{\noise}{\bZ}
\newcommand{\noisedev}{\sigma_{\!\ast}}
\newcommand{\noisevar}{\noisedev^{2}}
\newcommand{\pc}{P_{\!c}}
\newcommand{\pmax}{P_{\!\max}}
\newcommand{\rx}{N}
\newcommand{\tx}{M}

\DeclareMathOperator{\ee}{EE}

\DeclareMathOperator{\mirror}{\bG}
\newcommand{\rate}{R}

\newcommand{\herm}{\mathbb{H}}

\newcommand{\psd}[1]{\herm_{+}^{#1}}
\newcommand{\spectron}{\mathcal{D}}
\newcommand{\hessmat}{\bD}

\newcommand{\lyap}{H}

\newcommand{\db}{\textrm{dB}\xspace}
\newcommand{\dbm}{\textrm{dBm}\xspace}

\newcommand{\km}{\textrm{km}\xspace}
\newcommand{\kmh}{\textrm{km}/\textrm{h}\xspace}
\newcommand{\ms}{\textrm{ms}\xspace}

\newcommand{\hz}{\textrm{Hz}\xspace}
\newcommand{\khz}{\textrm{kHz}\xspace}
\newcommand{\mhz}{\textrm{MHz}\xspace}
\newcommand{\ghz}{\textrm{GHz}\xspace}

\newcommand{\acdef}[1]{\emph{\acl{#1}} \textup(\acs{#1}\textup)\acused{#1}}

\begin{document}

%*************************************************************
%*****    FRONT MATTER
%*************************************************************

%----------------------------------------------------------------------
%%% TITLE & AUTHORS
%----------------------------------------------------------------------
%\title{%
%Exponential Learning:
%Distributed Optimization in Signal Processing and Networks under Uncertainty}

\title{%
Distributed Stochastic Optimization via\\
Matrix Exponential Learning}

\author{
Panayotis Mertikopoulos%
,~\IEEEmembership{Member,~IEEE},
E.~Veronica Belmega%
,~\IEEEmembership{Member,~IEEE},
\\ Romain~Negrel,
and
Luca Sanguinetti%
,~\IEEEmembership{Senior Member,~IEEE}\vspace{-0.9cm}
\thanks{%
P.~Mertikopoulos is with the French National Center for Scientific Research (CNRS),
and the Laboratoire d'Informatique de Grenoble (LIG), F-38000, Grenoble, France. 
E.~V.~Belmega is with ETIS\,/\,ENSEA \textendash\ UCP \textendash\ CNRS, Cergy-Pontoise, France and Inria, Grenoble, France.
R.~Negrel is with GREYC, CNRS UMR 6072, ENSICAEN
Université de Caen Basse-Normandie, France. 
L.~Sanguinetti is with the University of Pisa, Dipartimento di Ingegneria dell'Informazione, Italy, and also with the Large Systems and Networks Group (LANEAS), CentraleSup\'elec, Universit\'e Paris-Saclay, Gif-sur-Yvette, France.}
\thanks{%
This research was supported by
the European Reseach Council under grant no. SG-305123-MORE,
%the European Commission in the framework of the FP7 Network of Excellence in Wireless COMmunications NEWCOM\# (contract no. 318306),
the French National Research Agency project
NETLEARN (ANR\textendash 13\textendash INFR\textendash 004),
and by ENSEA, Cergy-Pontoise, France.
%L. Sanguinetti was supported by the ERC Starting Grant 305123 MORE.
}
} % end author block

\maketitle

%----------------------------------------------------------------------
%%% ACRONYMS
%----------------------------------------------------------------------
\newacro{KL}{Kullback\textendash Leibler}
\newacro{ODE}{ordinary differential equation}
\newacro{NE}{Nash equilibrium}
\newacroplural{NE}[NE]{Nash equilibria}
\newacro{SE}{stable equilibrium}
\newacroplural{SE}[SE]{stable equilibria}
\newacro{XL}{exponential learning}
\newacro{DSC}{diagonal strict concavity}
\newacro{DC}{diagonal concavity}
\newacro{LMI}{linear matrix inequality}
\newacroplural{LMI}{linear matrix inequalities}
\newacro{BER}{bit error rate}
\newacro{OLE}{online learning}
\newacro{FM}{Foschini\textendash Miljanic}
\newacro{EE}{energy efficiency}
\newacro{AIMD}{additive increase, multiplicative decrease}
\newacro{5G}{fifth generation}
\newacro{SISO}{single-input and single-output}
\newacro{MIMO}{mul\-tiple-input and multiple-output}
\newacro{MUI}{multi-user interference-plus-noise}
\newacro{MAC}{medium access control}
\newacro{CSI}{channel state information}
\newacro{CSIT}{channel state information at the transmitter}
\newacro{BS}{base station}
\newacro{TDD}{time-division duplexing}
\newacro{CDMA}{code division multiple access}
\newacro{FDMA}{frequency division multiple access}
\newacro{DSL}{digital subscriber line}
\newacro{SIC}{successive interference cancellation}
\newacro{SUD}{single user decoding}
\newacro{SINR}{signal-to-interference-and-noise ratio}
\newacro{KKT}{Ka\-rush--Kuhn--Tuc\-ker}
\newacro{WF}{water-filling}
\newacro{IWF}{iterative water-filling}
\newacro{SWF}{simultaneous water-filling}
\newacro{iid}[i.i.d.]{independent and identically distributed}
\newacro{OFDMA}{orthogonal frequency-division multiple access}
\newacro{MXL}{matrix exponential learning}
\newacro{AMXL}[MXL-a]{asynchronous matrix exponential learning}
\newacro{EXL}[MXL-eig]{eigen-based exponential learning}
\newacro{FCC}{Federal Communications Commission}
\newacro{NTIA}{National Telecommunications and Information Administration}
\newacro{GAO}{General Accounting Office}
\newacro{QoE}{quality of experience}
\newacro{QoS}{quality of service}
\newacro{OFDM}{orthogonal frequency division multiplexing}
\newacro{MIMO-OFDM}{multiple-input multiple-output orthogonal frequency division multiplexing}
\newacro{EW}{exponential weight}
\newacro{OGA}{online gradient ascent}
\newacro{OMD}{online mirror descent}
\newacro{APT}{asymptotic pseudotrajectory}
\newacro{ICT}{information and communications technology}
\newacro{MSE}{mean squared error}
\newacro{EPA}{extended pedestrian A}
\newacro{EVA}{extended vehicular A}
\newacro{ETU}{extended typical urban}
\newacro{UL}{uplink}
\newacro{DL}{downlink}
\newacro{CCI}{co-channel interference}

%----------------------------------------------------------------------
%%% ABSTRACT
%----------------------------------------------------------------------
%\vspace{-2cm}
\begin{abstract}
%----------------------------------------------------------------------
%%% ABSTRACT
%----------------------------------------------------------------------
% !TEX root = ./Main.tex
%
%
%Motivated by applications in signal processing and wireless communications,
In this paper, we investigate a distributed learning scheme for a broad class of stochastic optimization problems and games that arise in signal processing and wireless communications.
The proposed algorithm relies on the method of \ac{MXL} and only requires locally computable gradient observations that are possibly imperfect and/or obsolete.
To analyze it, we introduce the notion of a stable \acl{NE} and we show that the algorithm is globally convergent to such equilibria \textendash\ or locally convergent when an equilibrium is only locally stable.
We also derive an explicit linear bound for the algorithm's convergence speed, which remains valid under measurement errors and uncertainty of arbitrarily high variance.
To validate our theoretical analysis, we test the algorithm in realistic multi-carrier/multiple-antenna wireless scenarios where several users seek to maximize their \acl{EE}.
Our results show that learning allows users to attain a net increase
%the system reaches a \acl{NE} with a net increase
between $100\%$ and $500\%$ in \acl{EE}, even under very high uncertainty.
\end{abstract}

%----------------------------------------------------------------------
%%% KEYWORDS
%----------------------------------------------------------------------
\begin{IEEEkeywords}
Learning, stochastic optimization, game theory, matrix exponential learning, variational stability, uncertainty.
\end{IEEEkeywords}

%*************************************************************
%*****    BODY TEXT
%*************************************************************
\acresetall

%----------------------------------------------------------------------
%%% INTRODUCTION
%----------------------------------------------------------------------
\section{Introduction}
\label{sec:intro}
%----------------------------------------------------------------------
%%% INTRODUCTION
%----------------------------------------------------------------------
% !TEX root = ./Main.tex

\IEEEPARstart{C}{onsider} a finite set of optimizing \emph{players} (or \emph{agents}) $\playset = \{1,\dotsc,\players\}$, each controlling a positive-semidefinite matrix variable $\bX_{\play}$, and behaving selfishly so as to improve their individual well-being.
%as quantified by an associated \emph{utility} (or \emph{payoff}) \emph{function} $\pay_{\play}(\bX_{1},\dotsc,\bX_{\players})$.
%Denote $\bX_{-\play} = \left(\bX_{1},\ldots,\bX_{\play-1},\bX_{\play+1},\ldots\bX_{\play}\right)$ and consider the following set $\game$ of coupled optimization problems (in maximization form):
Assuming that this well-being is quantified by a \emph{utility} (or \emph{payoff}) \emph{function} $\pay_{\play}(\bX_{1},\dotsc,\bX_{\players})$, we obtain the coupled semidefinite optimization problem
\begin{flalign}
\label{eq:game}
\text{for all $\play\in\playset$}
	\quad
	\begin{cases}
	\textrm{maximize}
		&\quad
		\pay_{\play} (\bX_{1},\dotsc,\bX_{\players})
		\\
	\textrm{subject to}
		&\quad
		\bX_{\play} \in \feas_{\play}
	\end{cases}
\end{flalign}
where
$\feas_{\play}$ denotes the set of feasible actions of player $\play$.
%and
%we used the standard shorthand $(\bX_{\play};\bX_{-\play}) \equiv (\bX_{1},\dotsc,\bX_{\play},\dotsc,\bX_{\players})$.
Specifically, we will focus on feasible action sets of the general form
\begin{equation}
\label{eq:feasible}
\feas_{\play}
	= \setdef{\bX_{\play}\mgeq 0}{\norm{\bX_{\play}} \leq A_{\play}}
\end{equation}
where $\norm{\bX_{\play}} = \sum_{m=1}^{\tx} \abs{\mathrm{eig}_{m}(\bX_{\play})}$ denotes the nuclear matrix (or trace) norm of $\bX_{\play}$,
$A_{i}$ is a positive constant,
and
the players' utility functions $\pay_{\play}$ are assumed individually concave and smooth in $\bX_{\play}$ for all $\play\in\playset$.

The coupled multi-agent, multi-objective problem \eqref{eq:game} constitutes a \emph{game},
which we denote by $\game$.
As we discuss in the next section, games and optimization problems of this type are extremely widespread in signal processing, wireless communications and information theory, especially in a stochastic framework where:
\begin{inparaenum}
[\itshape a\upshape)]
\item
the objective functions $\pay_{\play}$ are themselves expectations over an underlying random variable (see Ex.~\ref{sec:image} below);
and/or
\item
the feedback to the optimizers is subject to noise and/or measurement errors (Ex.~\ref{sec:EE}).
\end{inparaenum}
Accordingly, our main goal will be to provide a learning algorithm that converges to a suitable solution of $\game$, subject to the following desiderata:
\begin{enumerate}
[\upshape(\itshape i\upshape)]
\item
\emph{Distributedness}:
player updates are based on local information and measurements.
\item
\emph{Robustness}:
feedback and measurements may be subject to random errors, noise, and delays.
\item
\emph{Statelessness}:
players are oblivious to the overall state (or interaction structure) of the system.
\item
\emph{Flexibility}:
players can employ the algorithm in both static and ergodic environments.
\end{enumerate}

To achieve this, we build on the method of \acdef{MXL} that was recently introduced by the authors of \cite{MM16} in the context of throughput maximization in \ac{MIMO} systems.
%More specifically, we propose a distributed optimization algorithm based on the method of \ac{MXL} that was recently introduced by the authors of \cite{MBM12, MM16}.
In a nutshell, the main idea of the proposed method is that each player tracks the individual gradient of his utility function via an auxiliary score matrix, possibly subject to randomness and/or feedback imperfections.
The players' actions are then computed via an ``exponential projection'' step that maps these score matrices to the players' action spaces, and the process repeats.
%In so doing, the aggregation of the players' gradient information acts as a filter that weeds out the noise in the long run,
{Thus, building on the preliminary results of \cite{MM16} for \ac{MIMO} throughput maximization, we show here that}
\begin{inparaenum}
[\itshape a\upshape)]
\item
\ac{MXL} can be applied to a much broader class of (stochastic) optimization problems and games of the general form \eqref{eq:game};
\item
{the algorithm's convergence does not require the (restrictive) structure of a potential game \cite{MS96};}
\item
\ac{MXL} converges to equilibrium (locally or globally) under very mild assumptions for the noise and uncertainty surrounding the optimizers' decisions;
and
\item
we derive explicit estimates for the method's rate of convergence.
\end{inparaenum}
%allowing the algorithm to converge to equilibrium (locally or globally) in a large class of (stochastic) optimization problems and games of the general form \eqref{eq:game}.
%\VB{We may want to reinforce what is different with [1,2] a bit.  Suggestion: add at the end "in the general form \eqref{eq:game}, going beyond the throughput maximization problem in [1,2]." or something like.}
%\PM{Restructured, whaddya think?}
%\VB{Good!}

%\subsection{Motivation and contributions}
%Constrained optimization problems of the form in $\mathcal G$ may arise in connection with a wide range of problems in different applications and settings in signal processing and networks.
%
\subsection{Related work}
\label{sec:related}

The \ac{MXL} method studied in this paper has strong ties to matrix regularization \cite{KSST12} and mirror descent methods \cite{Nes09,SS11} for (online) convex optimization.
In particular, the important special case of real vector variables on the simplex (real diagonal $\bX_{\play}$ with constant trace) is closely related to the well-known \ac{EW} learning algorithm for multi-armed bandit problems \cite{Vov90}.
More recently, \ac{MXL} schemes were also proposed for single-user regret minimization in the context of online power control \cite{SMT15} and throughput maximization \cite{MB14} for dynamic \ac{MIMO} systems.
The goal there was to show that, in the presence of temporal variabilities, the long-run performance of \acl{MXL} matches that of the best fixed policy in hindsight, a property known as ``no regret'' \cite{SS11,CBL06}.
Nonetheless, in a multi-user, game-theoretic setting, a no-regret strategy may end up assigning positive weight \emph{only} to strictly dominated actions \cite{VZ13}.
As a result, (external) regret minimization is neither necessary nor sufficient to ensure convergence to a \acl{NE} (or other game-theoretic solution concept).

In \cite{MBM12,MM16,YN15}, a special case of \eqref{eq:game} was studied in the context of \ac{MIMO} throughput maximization in the presence of stochasticity, in both single-user \cite{YN15} and multiple access channels \cite{MM16,MBM12}.
%games that admit a concave potential function.
%(not necessarily in a matrix setting).
%(specifically, rate maximization in Gaussian \aclp{MAC}).
%In the latter case, the existence of a potential greatly facilitates the analysis by essentially reducing \eqref{eq:game} to a simple ``common interest'' game \cite{MS96}.
In both cases, the problem boils down to a semidefinite optimization problem, possibly distributed over several agents \cite{MM16}.
The existence of a single objective function greatly facilitates the analysis;
however, many cases of practical interest (such as the examples we discuss in the following section) cannot be modeled as problems of this type, so a potential-based analysis is often unsuitable for practical applications.

In this paper, we derive the convergence properties of \acl{MXL} in concave $\players$-player games, and we investigate the algorithm's long-term behavior under feedback errors and uncertainty.
Specifically, building on earlier work by Rosen \cite{Ros65}, we consider a broad class of games that satisfy a local \emph{variational stability} condition which ensures that \aclp{NE} are locally isolated.
%\textendash\ all games that admit a concave potential are members of this class.
In a series of recent papers, Scutari et al. used a stronger, global variant of this condition to establish the convergence of a class of Gauss\textendash Seidel, best-response methods,
%for solving variational inequalities,
and successfully applied these algorithms to a wide range of communication problems \textendash\ for a panoramic survey, see \cite{SFPP10} and references therein.
However, in the presence of noise and uncertainty, the convergence of best-response methods is often compromised:
as an example, in the case of throughput maximization with imperfect feedback, \acl{IWF} (a standard best-response scheme) \cite{YRBC04} fails to provide any perceptible gains over crude, uniform power allocation policies \cite{MM16}.
Thus, given that randomness, uncertainty and feedback imperfections are ubiquitous in practical systems, we focus throughout on attaining convergence results that are robust to learning impediments of this type.

\subsection{Main contributions and paper outline}
\label{sec:outline}
The main contribution of this work is the derivation of the convergence properties of \acl{MXL} for games played over bounded regions of positive-semidefinite matrices in the presence of feedback noise and randomness.
To put our theoretical analysis in context, we first discuss three examples from wireless networks and computer vision in Section \ref{sec:examples}.
More specifically, we illustrate
\begin{inparaenum}
[\itshape i\upshape)]
\item
a general game-theoretic framework for \emph{contention-based} \acdef{MAC};
\item
a stochastic optimization formulation for \emph{content-based image retrieval}
{(a key ``Big Data'' signal processing problem);}
%which is gaining a lot of momentum in signal processing for Big Data;
and
\item
the multi-agent problem of transmit \acdef{EE} maximization in multi-carrier \ac{MIMO} networks (a critical design feature of emerging green multi-cellular networks).
\end{inparaenum}

In Section \ref{sec:model}, we revisit our core game-theoretic framework, and we discuss the notion of a \emph{stable} \acl{NE}.
%together with a set of easily verifiable sufficient conditions to check equilibrium stability.
Subsequently, in Section \ref{sec:learning}, we introduce our \acl{MXL} scheme, and we detail our assumptions for the stochasticity affecting the players' objectives and observations.
Our main results are then presented in Section \ref{sec:analysis} and can be summarized as follows:
Under fairly mild assumptions on the underlying stochasticity (zero-mean feedback errors with finite conditional variance), we show that
\begin{inparaenum}
[\upshape(\itshape i\hspace*{1pt}\upshape)]
\item
the algorithm's only termination states are \aclp{NE};
\item
if the game admits a globally (locally) stable equilibrium, then the algorithm converges globally (locally) to said equilibrium;
and
\item
on average, the algorithm converges to an $\eps$-neighborhood of an extreme (resp. interior) strongly stable equilibrium within $\bigoh(1/\eps)$ iterations (resp. $\bigoh(1/\eps^{2})$).
\end{inparaenum}
%Finally, we discuss some implementation variants of the proposed algorithm in order to
%\added{mitigate the impact of floating point overflows
%%floating point arithmetic errors
%and asynchronous/delayed agent updates.}

The above results greatly extend and generalize the recent analysis of \cite{MM16} for throughput maximization in \ac{MIMO} \ac{MAC} systems.
To further validate our results in \ac{MIMO} environments, our theoretical analysis is supplemented with extensive numerical simulations for \acl{EE} maximization in multi-carrier \ac{MIMO} wireless networks in Section \ref{sec:numerics}.
To streamline the flow of the paper, technical proofs have been relegated to a series of appendices at the end.

\smallskip
\subsubsection*{Notation}
{
In what follows, the profile $\bX = (\bX_{1},\dotsc,\bX_{\players})$ is identified with the block-diagonal matrix $\diag(\bX_{\play})_{\play=1}^{\players}$, and we use the game-theoretic shorthand $(\bX_{\play};\bX_{-\play})$ when we need to focus on the action $\bX_{\play}$ of player $\play$ against that of all other players.
Also, given $\bA\in\CC^{d\times d}$, we write $\norm{\bA} = \sum_{j=1}^{d} \abs{\mathrm{eig}_{j}(\bA)}$ for the nuclear (trace) norm of $\bA$ and $\dnorm{\bA} = \max_{j} \abs{\mathrm{eig}_{j}(\bA)}$ for its (dual) singular norm.
}
%$\herm^{n}$ for the space of $n\times n$ Hermitian matrices and $\herm_{+}^{n}$ for the positive-semidefinite cone of $\herm^{n}$.
%Finally, given $\bX\in\herm^{n}$, we will write $\|\bX\| = \sum_{j=1}^{n} |\lambda_{j}(\bX)|$ for the nuclear norm of $\bX$ and $\|\bX\|_{\infty} = \max_{j}|\lambda_{j}(\bX)|$ for its (dual) singular norm.

%----------------------------------------------------------------------
%%% EXAMPLES
%----------------------------------------------------------------------
\section{Motivation and Examples}
\label{sec:examples}
%----------------------------------------------------------------------
%%% EXAMPLE
%----------------------------------------------------------------------
% !TEX root = ./Main.tex

\newcommand{\dataset}{\mathcal{D}}
\newcommand{\sample}{\mathcal{W}}
\newcommand{\similar}{\mathcal{S}}
\newcommand{\dissimilar}{\mathcal{U}}
\newcommand{\triples}{\mathcal{T}}

To motivate the general framework of \eqref{eq:game}, we illustrate below three examples taken from communication networks and computer vision (and reflecting the subjective tastes and interests of the authors).
A reader who is interested in the general theory may skip this section and proceed directly to Sections \ref{sec:model}\textendash\ref{sec:analysis}.

The first example below defines a game-theoretic model for the interactions between wireless users in networks with contention-based medium access.
The second example revolves around metric learning for similarity-based image search and showcases the range of problems where the proposed method applies.
Finally, the third example focuses on \acf{EE} maximization in \ac{MIMO} multi-user networks, a key aspect of future and emerging wireless networks:
to the best of our knowledge, \ac{MXL} provides the first distributed and provably convergent algorithm for \acl{EE} maximization in with noisy feedback and imperfect \acl{CSI}.

%----------------------------------------------------------------------
%%% CONTENTION
%----------------------------------------------------------------------
\subsection{Contention-based medium access}

Contention-based \acf{MAC} aims to provide an efficient means for accessing and sharing a wireless channel in the presence of several competing wireless users that interfere with each other.
To model this, consider a set of wireless users indexed by $\playset = \{1,\dotsc,\players\}$,
each updating their individual channel access probability $x_{\play}$ in response to the amount of contention in the network \cite{Chen2010}.
In practice, wireless users cannot be assumed to know the exact channel access probabilities of other users, so user $i$ infers the level of wireless contention via an aggregate \emph{contention measure} $q_{\play}(\bx_{-\play})$ which is determined as a (symmetric) function of the access probability profile $\bx_{-\play} = (x_{1},\dotsc,x_{i-1},x_{i+1}\dotsc,x_{\players})$ of all other users.%
\footnote{For instance, a standard contention measure of this type is the conditional collision probability
$q_{\play}({\bf x}_{-\play}) = 1 - \prod_{\playalt\neq\play} (1-x_{\playalt})$. \cite{Chen2010}.}

With this in mind, the objective of each user is to select their individual channel access probability so as to maximize the benefit derived from acessing the channel more often minus the induced contention $x_{\play} q_{\play}(\bx_{-\play})$ incurred by all other users.
This leads to the utility function formulation
\begin{equation}
\pay_{\play}(x_{\play};\bx_{-\play})
	= U_{\play}(x_{\play}) - x_{\play} q_{\play}(\bx_{-\play})
\end{equation}
where $U_{\play}$ is a continuous and nondecreasing function representing the utility of user $\play$ when there are no other users in the channel.
Thus, in economic terms, $\pay_{\play}$ simply represents the user's net gain from channel access, discounted by the associated contention cost.

We thus obtain the multi-agent, multi-objective formulation
\begin{flalign}
\label{eq:game-MAC}
\text{for all $\play\in\playset$}
	\quad
	\begin{cases}
	\textrm{maximize}
		&\quad
		\pay_{\play} (x_{\play};\bx_{-\play}),
		\\
	\textrm{subject to}
		&\quad
		x_{\play} \in [0,1].
	\end{cases}
\end{flalign}
whose solutions can be analyzed through the specification of the utility function $U_{i}(x_{i})$ and the choice of the contention measure $q_{i}$. If $U_{i}(x_{i})$ is assumed to be continuously differentiable and strictly concave (see for example \cite{Cui08,Chen2010} and references therein), \eqref{eq:game-MAC} is a special case of \eqref{eq:game} with $M=1$ and $A_{\play}=1$.
We further note the unilateral optimization problems above are interconnected because of the dependence of the contention measure $q_{i}(\bx_{-i})$ on the access probabilities of all users;
as a result, the theory of noncooperative games theory arises naturally as the most suitable framework to analyze and solve \eqref{eq:game-MAC} \cite{Cui08}.

%----------------------------------------------------------------------
%%% METRIC LEARNING
%----------------------------------------------------------------------
\subsection{Metric learning for image similarity search}
\label{sec:image}

%\PM{To be done (Luca?): insist somewhere that this is an optimization problem, not a game.}
%\LS{Can't we think of any application in which it might be a distributed system? I was thinking about radar networks for surveillance or facial recognition systems with multiple cameras. They might be distributed systems (with no central unit - I guess) but it also true that if there is no competition what is the point to make this exercise. What else? Take it as my last observation! But no big deal, as it is, it fine for me! :)}

A key challenge in content-based image retrieval is to design an automatic procedure capable of retrieving documents from a large database based on their similarity to a given request, often distributed over several computing cores in a massively parallel computing grid (or cloud) \cite{Negrel-2013,Smeulders-2000}.
To formalize this, an image is typically represented via its \emph{signature}, i.e. a real $d$-dimensional vector $\bi\in\R^{d}$ that collects and encodes the most distinctive features of said image.
Given a database $\dataset$ of such signatures, each image $\bi\in\dataset$ is further associated with a set $\similar_{\bi}\subseteq\dataset$ of \emph{similar} images and a set $\dissimilar_{\bi}\subseteq\dataset$ of \emph{dissimilar} ones, based on each image's content.
Accordingly, the goal is to design a distance metric which is minimized between similar images and is maximized between dissimilar ones.
%in so doing, image retrieval can be automated by simply searching for database images that are closest to the given request.

A widely used distance measure between image signatures is the \emph{Mahalanobis distance}, defined as
\begin{equation}
\label{eq:Mahalanobis}
d_{\bX} (\bi,\bj) = (\bi-\bj)^{\top}\, \bX \, (\bi - \bj),
\end{equation}
where $\bX \mgeq 0$ is a $d\times d$ positive-definite matrix.%
\footnote{The baseline case $\bX = \bI$ corresponds to the Euclidean distance, which is often unsuitable for image discrimination purposes \cite{Bellet-2013}.}
This so-called ``precision matrix'' must then be chosen by the optimizer so that $d_{\bX} (\bi,\bj) < d_{\bX} (\bi,\bk)$ whenever $\bi$ is similar to $\bj$ but dissimilar to $\bk$.
With this in mind, we obtain the minimization objective
\begin{equation}
\label{eq:metric-all}
F(\bX;\triples)
	= \sum_{\mathclap{(\bi, \bj,\bk) \in\triples}} C\big(d_{\bX}(\bi,\bj) -  d_{\bX}(\bi,\bk) - \eps\big)
	+ \norm{\bX - \bI}_{F}^{2},
\end{equation}
where:
\begin{inparaenum}
[\itshape a\hspace*{1pt}\upshape)]
\item
$\triples$ is the set of all triples $(\bi,\bj,\bk)$ such that $\bj\in\similar_{\bi}$ and $\bk\in\dissimilar_{\bi}$;
\item
$C$ is a convex, nondecreasing function that penalizes matrices $\bX$ that do not capture similarities and dissimilarities between images;
\item
the parameter $\eps > 0$ reinforces this penalty;
and
\item
$\norm{\cdot}_{F}$ denotes the ordinary Frobenius ($\ell_{2}$) norm.%
\footnote{This last regularization term is included in order to maximize predictive accuracy by reducing the effects of over-fitting to training data \cite{Law-2014}.}
\end{inparaenum}

An additional requirement in the above is to employ a low-rank precision matrix $\bX$ so as to
reduce model complexity and computational costs \cite{Negrel-2013},
enable distributed storing and retrieval \cite{Lim-2013},
and better exploit correlations between features that further reduce over-fitting effects \cite{Law-2014}.
A computationally efficient way to achieve this is to include a trace constraint of the form $\tr\bX \leq c$ for some $c\ll d$,%
\footnote{By contrast, an $\ell_{0}$ rank constraint of the form $\rank(\bX) \leq c$ generically leads to an untractable NP-hard problem formulation.}
leading to the feasible region
\begin{equation}
\label{eq:metric-feasible}
\feas
	= \setdef{\bX\in\R^{d\times d}}{\bX \mgeq 0 \;\text{and}\; \tr\bX \leq c}.
\end{equation}

Combining \eqref{eq:metric-all} and \eqref{eq:metric-feasible}, we see that metric learning in image retrieval is a special case of the general problem \eqref{eq:game} with $\players=1$ optimizers.
%and real positive-definite matrix variables instead of complex ones.
However, when $\dataset$ is large, the number of variables involved is computationally prohibitive:
for instance, \eqref{eq:metric-all} may contain up to $10^{9}$\textendash$10^{11}$ terms for a modest database with $10^{4}$ images.
To circumvent this obstacle, a common approach is to replace $\triples$ with a smaller, randomly drawn population sample $\sample\subseteq\triples$, and then take the average over all such samples.
In so doing, we obtain the stochastic optimization problem \cite{Bottou-2012}:
\begin{equation}
\label{eq:metric-stochastic-optim}
\begin{aligned}
\textrm{minimize}
	&\quad
%	u(\bX) \equiv
	\exof{F(\bX; \sample)}
	\hspace{1ex}
	%\textrm{for all $\play = 1,\dotsc,\plays$,}
	\\
\textrm{subject to}
	&\quad
	\bX \mgeq 0, \tr(\bX) \leq c,
\end{aligned}
\end{equation} 
where the expectation is taken over the random samples $\sample$.

The benefit of this formulation is that, at each realization, only a small-size, tractable data sample $\sample$ is used for calculations at each computing node.
On the flip side however, the expectation in \eqref{eq:metric-stochastic-optim} cannot be calculated, so the optimizer only has access to information on the random, realized gradients $\nabla_{\bX} F(\bX;\sample)$.
This type of uncertainty is typical of stochastic optimization problems, and the proposed \ac{MXL} method has been designed precisely with this feedback structure in mind.

%----------------------------------------------------------------------
%%% METRIC LEARNING
%----------------------------------------------------------------------
\subsection{Energy efficiency maximization}
\label{sec:EE}

Consider the problem of \acl{EE} maximization in multi-user, multiple-carrier \ac{MIMO} networks \textendash\ see e.g. \cite{BSHD15,ZSBJD16,MB16} and references therein.
Here, wireless connections are established over transceiver pairs with $\tx$ (resp. $\rx$) antennas at the transmitter (resp. receiver), and communication takes place over a set of $\carriers$ orthogonal subcarriers.
Accordingly, the $\play$-th transmitter is assumed to control his individual input signal covariance matrix $\bQ_{\play\carrier}$ over each subcarrier $\carrier=1,\dotsc,\carriers$, subject to the constraints:
\begin{inparaenum}
[\itshape a\upshape)]
\item
$\bQ_{\play\carrier} \mgeq 0$ (since each $\bQ_{\play\carrier}$ is a signal covariance matrix);
and
\item
$\tr\bQ_{\play} \leq \pmax$,
where
$\bQ_{\play} = \diag(\bQ_{\play\carrier})_{\carrier=1}^{\carriers}$ is the covariance profile of the $\play$-th transmitter,
$\tr(\bQ_{\play})$ represents the user's transmit power over all subcarriers,
and
$\pmax$ denotes the user's \emph{maximum} transmit power.
\end{inparaenum}

\begin{figure}[t]
\centering
\input{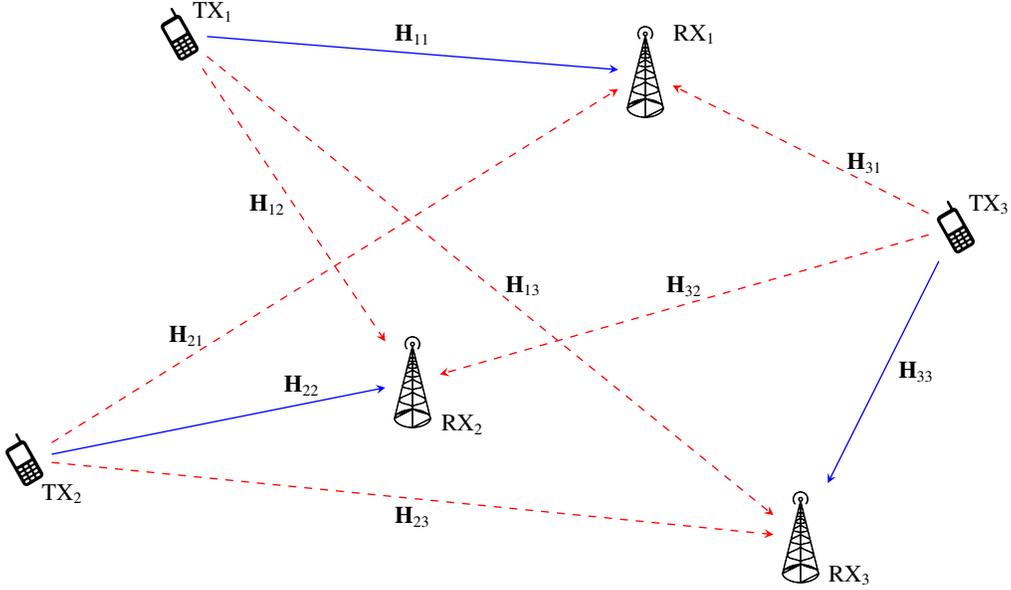}
\caption{A multi-user \ac{MIMO} system with $\plays=3$ connections.}
\vspace{-2ex}
\end{figure}

Assuming Gaussian input and \ac{SUD} at the receiver, each user's Shannon-achievable throughput is given by the well-known expression
\begin{equation}
\label{eq:rate}
\rate_{\play}(\bQ)
	= \log\det\big( \bW_{-\play} + \bH_{\play\play}\bQ_{\play}\bH_{\play\play}^{\dag} \big)
	- \log\det\big( \bW_{-\play} \big),
\end{equation}
where
$\bH_{\playalt\play}$ denotes the channel matrix profile between the $\playalt$-th transmitter and the $\play$-th receiver over all subcarriers,
$\bQ = (\bQ_{1},\dotsc,\bQ_{\players})$ denotes the users' transmit profile and
$\bW_{-\play}
	\equiv \bW_{-\play}(\bQ)
	= \bI + \insum_{\playalt\neq\play} \bH_{\playalt\play} \bQ_{\playalt} \bH_{\playalt\play}^{\dag}$
is the \ac{MUI} covariance matrix at receiver $\play$.
The users' transmit \acdef{EE} is then defined as the ratio between their Shannon rate and the total consumed power, i.e.
\begin{equation}
\label{eq:EE}
\ee_{\play}(\bQ)
%	= \frac{\rate_{\play}(\bQ)}{\pc + P_{\play}}
	= \frac{\rate_{\play}(\bQ)}{\pc + \tr(\bQ_{\play})},
\end{equation}
where $\pc>0$ represents the total power consumed by circuit components at the transmitter \cite{BSHD15}.

The energy efficiency function above is not concave w.r.t the covariance matrix $\bQ_{\play}$ of user $i$,
but it can be recast as such via a suitable Charnes-Cooper transformation \cite{CC62}.
To this aim, following \cite{MB16}, consider the adjusted control variables
\begin{equation}
\label{eq:Xdef}
\bX_{\play}
	= \frac{\pc + \pmax}{\pmax} \frac{\bQ_{\play}}{\pc + \tr(\bQ_{\play})},
\end{equation}
where the normalization constant $(\pc + \pmax)/\pmax$ implies that $\tr(\bX_{\play}) \leq 1$ with equality if and only if $\tr(\bQ) = \pmax$.
The action set of user $\play$ is thus given by
\begin{equation}
\label{eq:new_strat}
\txs
\feas_{\play}
	= \setdef{\diag(\bX_{\play\carrier})_{\carrier=1}^{\carriers}}{\bX_{\play\carrier} \mgeq0 \text{ and } \insum_{\carrier} \tr \bX_{\play\carrier} \leq 1},
\end{equation}
and, using \eqref{eq:Xdef},
the \acl{EE} expression \eqref{eq:EE} yields the utility function
\begin{align}
\label{eq:new_EE}
\pay_{\play}(\bX_{i};\bX_{-i})
	= \frac{ \pc + (1 - \tr\bX_{\play})\,\pmax}{\pc(\pc + \pmax)}
	\log\det\left({
	\bI + \frac{ \pc \pmax\, \effH_{\play} \bX_{\play} \effH_{\play}^{\dag}}{\pc + (1 - \tr\bX_{\play}) \pmax}
	}\right),
\end{align}
where $\effH_{\play}  = \cramped{\bW_{-\play}^{-1/2}} \bH_{\play \play}$ denotes the \emph{effective channel matrix} of user $\play$.

We thus see that the resulting \acl{EE} maximization problem is of the general form \eqref{eq:game} with $A_{\play} = 1$.%
\footnote{Strictly speaking, the block-diagonal constraint in \eqref{eq:new_strat} does not appear in \eqref{eq:feasible}, but it is satisfied automatically by the learning method presented in Sec.~\ref{sec:learning}.}
In this context, randomness and uncertainty stem from the noisy estimation of the users' \ac{MUI} covariance matrices (which depend on the other users' behavior),
the noise being due to the scarcity of perfect \ac{CSIT}, random measurement errors, etc.
To the best of our knowledge, the \ac{MXL} method discussed in Section \ref{sec:learning}
comprises the first distributed solution scheme for \acl{EE} maximization in general multi-user/multi-antenna/multi-carrier networks with local, causal \textendash\ and possibly imperfect \textendash\ \ac{CSI} feedback at the transmitter.

%----------------------------------------------------------------------
%%% GAME THEORY
%----------------------------------------------------------------------
\section{Elements from Game Theory}
\label{sec:model}
%----------------------------------------------------------------------
%%% MODEL
%----------------------------------------------------------------------
% !TEX root = ./Main.tex

The most widely used solution concept in noncooperative games is that of a \acdef{NE}, defined as any action profile $\eq\in\feas$ which is \emph{unilaterally stable} in the sense that
\begin{equation}
\label{eq:Nash}
%\tag{NE}
\pay_{\play}(\eq)
	\geq \pay_{\play}(\bX_{\play};\eq_{-\play})
	\quad
	\text{for all $\bX_{\play}\in\feas_{\play}$, $\play\in\playset$}.
\end{equation}
In other words, $\eq\in\feas$ is a \acl{NE} when no single player can further increase his individual utility assuming that the actions of all other players remain unchanged.

In complete generality, a game need not admit an equilibrium.
However, thanks to the concavity of each player's payoff function $\pay_{\play}$ and the compactness of their action space $\feas_{\play}$, the existence of a \acl{NE} in the case of \eqref{eq:game} is guaranteed by the general theory of \cite{Deb52}.
Hence, a natural question that arises is whether such an equilibrium solution is unique or not.

To answer this question, Rosen \cite{Ros65} provided a first-order sufficient condition for general $\players$-person concave games known as \acdef{DSC}.
To state it, define the \emph{individual payoff gradient} of player $\play$ as
\begin{equation}
\label{eq:gradient}
\payv_{\play}(\bX)
%	= \grad_{\play} \pay_{\play}(\bX)
	\equiv \grad_{\bX_{\play}} \pay_{\play}(\bX_{\play};\bX_{-\play}),
\end{equation}
and let $\payv(\bX) \equiv \diag(\payv_{1}(\bX),\dotsc,\payv_{\plays}(\bX))$ denote the collective profile of all players' individual gradients.
With this definition at hand, Rosen's condition can be stated as:
\begin{equation}
\label{eq:DSC}
\tag{DSC}
\trof{(\bX' - \bX)\big(\payv(\bX') - \payv(\bX)\big)}
	\leq 0
	\quad
	\text{for all $\bX,\bX'\in\feas$},
\end{equation}
with equality if and only if $\bX=\bX'$.
%\footnote{In the language of variational analysis, \eqref{eq:DSC} simply posits that $\payv(\bX)$ is a strictly monotone operator \cite{FP03}.}
%This condition was also examined in \cite{HS09,MS16b} with respect to the convergence of evolutionary dynamics in population games.
We then have:

\begin{theorem}[Rosen \cite{Ros65}]
\label{thm:Rosen}
If a game of the general form \eqref{eq:game} satisfies \eqref{eq:DSC}, then it admits a unique \acl{NE}.
\end{theorem}

The above theorem provides a sufficient condition for equilibrium uniqueness, but it does not provide a way for players to compute it \textendash\ especially in a decentralized setting with no information exchange between players and/or imperfect feedback.
More recently, \eqref{eq:DSC} was used by Scutari et al. (see \cite{SFPP10} and references therein) as the starting point for the convergence analysis of a class of Gauss\textendash Seidel methods for concave games based on variational inequalities \cite{FP03}.
Our approach is similar in scope but relies instead on the following notion of stability:

\begin{definition}
\label{def:stability}
The profile $\eq\in\feas$ is called \emph{stable} if it satisfies the variational stability condition:
\begin{equation}
\label{eq:VS}
\tag{VS}
\trof{(\bX - \eq) \, \payv(\bX)}
	\leq 0 
	\;\;
	\text{for all $\bX$ sufficiently close to $\eq$}.
\end{equation}
In particular, if \eqref{eq:VS} holds for all $\bX\in\feas$, we say that $\eq$ is \emph{globally stable}.
\end{definition}

%\begin{definition}[Local stability]
%\label{def:stablility}
%The profile $\eq\in\feas$ is called \emph{locally stable} if the following \emph{variational stability} condition
%\begin{equation}
%\label{eq:VS}
%\trof{(\bX - \eq) \, \payv(\bX)}
%	\leq 0
%\end{equation}
%\end{definition}
%
%\begin{definition}[Global stability]
%\label{def:stablility}
%The profile $\eq\in\feas$ is called \emph{globally stable} if \eqref{eq:VS} holds $\forall \bX\in\feas$.
%\end{definition}
%
%\textcolor{red}{[If you want, we can briefly comment the fact that \eqref{eq:VS} is less stringent of \eqref{eq:DSC} and gives some intuition for the condition itself. I think that this is what Veronica was asking for. We could also refer to the remark below on potential games.]}
%
%\textcolor{blue}{[PM: I included a comment to that effect. That said, we have other things we want to say, so I kept things short.]}

Mathematically,
\eqref{eq:VS} is implied by \eqref{eq:DSC};%
\footnote{Simply note that Nash equilibria are solutions of the variational inequality $\trof{(\bX - \eq) \payv(\eq)} \leq 0$ \cite{SFPP10}. Then, \eqref{eq:VS} follows by setting $\eq = \bX'$ in \eqref{eq:DSC}.}
the converse however does not hold, even when \eqref{eq:VS} holds globally.
Still, as is the case with \ac{DSC}, stability plays a key role in the characterization of \aclp{NE}:

\begin{proposition}
\label{prop:Nash-stable}
If $\eq\in\feas$ is stable, then it is an isolated \acl{NE};
specifically, if $\eq$ is globally stable, it is the game's unique \acl{NE}.
\end{proposition}

\begin{IEEEproof}
Suppose $\eq$ is stable, pick some $\bX_{\play}$ close to $\eq_{\play}$, and let $\bX = (\bX_{\play};\eq_{-\play})$.
Then, by \eqref{eq:VS}, we get $\trof{(\bX_{\play} - \eq_{\play}) \, \payv_{\play}(\bX_{\play};\eq_{-\play})} < 0$, implying that $\pay_{\play}$ is decreasing along the ray $\eq_{\play} + t (\bX_{\play} - \eq_{\play})$.
Since this covers all rays starting at $\eq_{\play}$, we conclude that $\eq$ is the game's unique equilibrium in the neighborhood of $\eq$ where \eqref{eq:VS} holds.
Our assertion then follows immediately.
\end{IEEEproof}

\smallskip

The variational stability condition \eqref{eq:VS} is important not only for characterizing the structure of the game's Nash set, but also for determining the convergence properties of the proposed learning scheme.
More precisely, as we show in Section \ref{sec:analysis},
local stability implies that a \acl{NE} is locally attracting, while global stability implies that it is globally so.
%if the game admits a globally (resp. locally) stable equilibrium, then the proposed learning algorithm converges globally (resp. locally) to said equilibrium.
Therefore, it is of paramount importance to have a verifiable criterion for the stability of a \acl{NE}. This can be accomplished by appealing to a second-order condition (similar to the second derivative test in calculus).

Specifically, define the \emph{Hessian} of a game as follows:
\begin{definition}
\label{def:Hessian}
The \emph{Hessian matrix} of a game is the block matrix $\hessmat(\bX) = \parens[\big]{\hessmat_{\play\playalt}(\bX)}_{\play,\playalt\in\playset}$ with blocks
\begin{equation}
\label{eq:Hessian}
\hessmat_{\play\playalt}(\bX)
	= \tfrac{1}{2}\nabla_{\bX_{\play}} \nabla_{\bX_{\playalt}} \pay_{\playalt}(\bX)
	+ \tfrac{1}{2} \bracks[\big]{\nabla_{\bX_{\playalt}} \nabla_{\bX_{\play}} \pay_{\play}(\bX)}^{\dag}.
%	= \nabla_{\bX_{\playalt}} \payv_{\play}(\bX)
%	+ \nabla_{\bX_{\play}} \payv_{\playalt}(\bX)^{\dag},\;
%	\play,\playalt\in\playset.
\end{equation}
%provided that the payoff functions of $\game$ are sufficiently smooth.
\end{definition}
%By definition, $\hessmat$ is block-Hermitian, i.e. $\hessmat_{\playalt\play} = \hessmat_{\play\playalt}^{\dag}$ for all $\playalt,\play \in \playset$;
The terminology ``Hessian'' reflects the fact that, in the single-player case where \eqref{eq:game} is a single-agent optimization problem, $\hessmat(\bX)$ is simply the Hessian of the optimizer's objective.
Thus, just as negative-definiteness of the Hessian of a function guarantees (strong) concavity and the existence of a unique maximizer, we have:

\begin{proposition}
\label{prop:Hessian}
If $\hessmat(\bX) \ml 0$ for all $\bX\in\feas$,
the game admits a unique and globally stable \acl{NE}.
More generally, if $\eq$ is a \acl{NE} and $\hessmat(\eq) \ml 0$, $\eq$ is locally stable and isolated.
\end{proposition}

\begin{IEEEproof}
The first statement of the proposition can be proved as follows. Assume first that $\hessmat(\bX) \ml 0$ for all $\bX\in\feas$.
Then, from \cite[Theorem 6]{Ros65} it follows that that the game satisfies \eqref{eq:DSC} and thus admits a unique \acl{NE} $\eq$; since 
\eqref{eq:DSC} implies \eqref{eq:VS} for all $\bX\in\feas$, our claim is immediate.
The proof for the second statement is as follows. If $\eq$ is a \acl{NE} and $\hessmat(\eq) \ml 0$, we will have $\hessmat(\bX) \ml 0$ for all $\bX$ in some convex neighborhood $\mathcal U$ of $\eq$ (by continuity).
%Without loss of generality, we may take $U$ to be convex;
Then, by applying \cite[Theorem 6]{Ros65} to $\mathcal U$ and reasoning as in the global case above, it follows that $\eq$ is the game's unique equilibrium in $\mathcal U$, as claimed.
\end{IEEEproof}

\begin{remark}
 In the special class of potential games, the players' payoff functions are aligned along a common goal, the game's potential \cite{MS96};
%i.e., to optimize a concave potential (or objective) function, all local maximizers are also global maximizers.
as a result, the maximizers of the game's potential are Nash equilibria.
%and if the potential function is concave, the converse is also true.
In this context, \eqref{eq:DSC} boils down to strict concavity of the potential function, which in turn implies the existence of a unique \acl{NE}.
Similarly,
%\eqref{eq:VS} reduces to the optimality condition of the maximum points.
the sufficient condition of Proposition \ref{prop:Hessian} reduces to the second order condition of concave potential functions that guarantees uniqueness of the solution (strictly negative definite Hessian matrix). 
\end{remark}

%----------------------------------------------------------------------
%%% LEARNING
%----------------------------------------------------------------------
\section{Learning under Uncertainty}
\label{sec:learning}
%----------------------------------------------------------------------
%%% LEARNING
%----------------------------------------------------------------------
% !TEX root = ./Main.tex

The goal of this section is to provide a learning algorithm that allows players to converge to a \acl{NE} in a decentralized environment with no information exchange between players and/or imperfect feedback.
To that end,
%motivated by the recent results of \cite{MBM12,MM16} for throughput maximization in \ac{MIMO} systems,
the main idea of the proposed learning scheme is as follows:
First, at every stage $n = 0,1,\dotsc$ of the process, each player $\play\in\playset$ tracks the individual gradient of his utility function via an auxiliary ``score'' matrix $\bY_{\play}(n)$, possibly subject to feedback/stochasticity-induced errors.
Subsequently, the players' actions $\bX_{\play}(n)$ are computed via an ``exponential projection'' step that maps the gradient tracking matrix $\bY_{\play}(n)$ back to the player's action space $\feas_{i}$, and the process repeats.

More precisely, we will focus on the following \acdef{MXL} scheme (for a pseudocode implementation, see Algorithm \ref{alg:MXL}):
\begin{equation}
\label{eq:MXL}
\tag{MXL}
\begin{aligned}
\bY_{\play}(n+1)
	&= \bY_{\play}(n) + \step_{n} \hat\payv_{\play}(n),
	\\
\bX_{\play}(n+1)
	&= A_{\play} \frac{\exp(\bY_{\play}(n+1))}{1 + \norm{\exp(\bY_{\play}(n+1))}},
\end{aligned}
\end{equation}
where:
\begin{enumerate}
\item
$n = 0,1,\dotsc$ denotes the stage of the process.
\item
the auxiliary matrix variables $\bY_{\play}(n)$ are initialized to an arbitrary (Hermitian) value.
\item
$\hat\payv_{\play}(n)$ is a stochastic estimate of the individual gradient $\payv_{\play}(\bX(n))$ of player $\play$ at stage $n$ (more on this below).
\item
$\step_{n}$ is a decreasing step-size sequence, typically of the form $\step_{n} \sim 1/n^{a}$ for some $a\in(0,1]$.
%\item
%$A_{\play} \in [0,\infty)$ represents the trace constraint of each player \textendash\ recall the definition \eqref{eq:feasible} of the players' action sets.
\end{enumerate}

\begin{algorithm}[t]
{%
\small
\sf
%\vspace{2pt}
\textbf{Parameter:}\;
%offset $\bias_{\play}>0$,
step-size sequence $\step_{n} \sim 1/n^{a}$, $a\in(0,1]$.
\\[2pt]
\textbf{Initialization:}\;
	$n \leftarrow 0$;\;
	$\bY_{\play} \leftarrow \text{any $\tx_{\play}\times\tx_{\play}$ Hermitian matrix}$.
	\\[2pt]
\Repeat
{%
%$\step\leftarrow \step_{n}$;
%\\[2pt]
	$n \leftarrow n+1$;
	\\[2pt]
	\ForEach
		{player $\play\in\playset$}
		{%
		play\;
		$\dis \bX_{\play} \leftarrow A_{\play} \frac{\exp(\bY_{\play})}{1 + \norm{\exp(\bY_{\play})}}$;
		\\[2pt]
		get gradient feedback $\hat\payv_{\play}$;
		\\[2pt]
		update auxiliary matrix\;
		$\bY_{\play} \leftarrow \bY_{\play} + \step_{n} \hat\payv_{\play}$;
%		\frac{\exp(\bY_{\play})}{\tr[\exp(\bY_{\play})]}$;
	} % end for
	\vspace{.5ex}
	\kwd{until} termination criterion is reached.
} % end Repeat
} % end \sf
\caption{Matrix exponential learning (\acs{MXL}).}
\acused{XL}
\label{alg:MXL}
\end{algorithm}

%\added{
%The recursion \eqref{eq:MXL} will be the centerpiece of our paper, so it is worth pointing out that it enjoys the following desirable properties (essentially by construction):
%\begin{enumerate}
%[\noindent (P1)]
%\item
%\emph{Distributedness:}
%players only require individual gradient information as feedback.
%\item
%\emph{Statelessness:}
%players do not need to know the state of the system.
%\item
%\emph{Robustness:}
%the players' feedback could be stochastic, imperfect, or otherwise perturbed by random noise.
%\item
%\emph{Reinforcement:}
%players tend to increase their individual utilities.
%\end{enumerate}
%}

Setting aside for a moment the precise nature of the stochastic estimates $\hat\payv_{\play}(n)$, we note that the update of the auxiliary matrix variable $\bY_{\play}(n)$ in \eqref{eq:MXL} acts as a ``steepest ascent'' step along the estimated direction of each player's individual payoff gradient $\payv_{\play}(\bX(n))$.%
\footnote{The step-size sequence $\step_{n}$ further finetunes this process and its role is discussed in detail later.}
As such, if there were no constraints for the players' actions, $\bY_{\play}(n)$ would define an admissible sequence of play and, \emph{ceteris paribus}, player $\play$ would tend to increase his payoff along this sequence.
However, this simple ascent scheme does not suffice in our constrained framework,
so $\bY_{\play}(n)$ is first exponentiated and subsequently normalized in order to meet the feasibility constraints \eqref{eq:feasible}.%
\footnote{Recall here that $\payv_{\play}$ is Hermitian as the derivative of a real function with respect to a Hermitian matrix variable \cite{Dat05}.
Moreover, the normalization step in \eqref{eq:MXL} subsequently ensures that the resulting matrix has $\norm{\bX_{\play}} \leq A_{\play}$, so the sequence $\bX_{\play}(n)$ induced by \eqref{eq:MXL} meets the problem's feasibility constraints.}

Of course, the outcome of the players' gradient tracking process depends crucially on the quality of the gradient feedback $\hat\payv_{\play}(n)$ that is available to them.
With this in mind, we will consider the following sources of uncertainty:
\begin{enumerate}
[\itshape i\upshape)]
\item
The players' gradient observations are subject to noise and/or measurement errors (cf. Ex.~\ref{sec:EE}).
\item
The players' utility functions are themselves stochastic expectations of the form $\pay_{\play}(\bX) = \exof{\hat\pay_{\play}(\bX;\omega)}$ for some random variable $\omega$, and the players can only observe the (stochastic) gradient of $\hat\pay_{\play}$ (cf. Ex.~\ref{sec:image}).
\item
Any combination of the above.
\end{enumerate}
%Similarly, in a stochastic optimization context, if the players' utility functions are themselves given by an expectation over a random variable (such as a channel's ergodic capacity in the presence of fast fading \cite{MM16}), the players might only be able to observe a random realization of their individual gradients \textendash\ and not their mean values.
%\textcolor{blue}{[PM: I think it would be ideal here if a) we could foreshadow this in the examples section; and b) we could tailor Example \ref{sec:access} to reflect the ``combination'' item above (since we can have both an expected utility and noise in the observations).]}

In view of all this, we will focus on the general model:
\begin{equation}
\label{eq:grad-noise}
\hat\payv_{\play}(n)
	= \payv_{\play}(\bX(n)) + \noise_{\play}(n),
\end{equation}
where the stochastic noise process $\noise(n)$ satisfies the hypotheses:
\begin{enumerate}
[({H}1)]

\item
\label{hyp:zeromean}
\emph{Zero-mean:}
\begin{equation}
\tag{H1}
\label{eq:zeromean}
\exof*{\noise(n) \given \bX(n)}
	= 0.
\end{equation}

\item
\label{hyp:MSE}
\emph{Finite \ac{MSE}:}
\begin{equation}
\tag{H2}
\label{eq:MSE}
\exof{\dnorm{\noise(n)}^{2} \given \bX(n)}
	\leq \noisevar
	\quad
	\text{for some $\noisedev>0$}.
\end{equation}
%for some $\noisedev>0$.
\end{enumerate}
The statistical hypotheses \eqref{eq:zeromean} and \eqref{eq:MSE} above are fairly mild from and allow for a broad range of estimation scenarios.%
\footnote{In particular, we will \emph{not} be assuming that the errors are \acs{iid} or bounded.}
%this observation is crucial for telecommunication systems because wireless feedback errors are typically correlated with the state of the network.}
In more detail, the zero-mean hypothesis \eqref{eq:zeromean} is a minimal requirement for feedback-driven systems, simply positing that there is no \emph{systematic} bias in the players' information.
Likewise, Hypothesis \eqref{eq:MSE} is a bare-bones assumption for the variance of the players' feedback, and it is satisfied by most common error processes \textendash\ such as Gaussian, log-normal, uniform and all sub-Gaussian distributions.
%controls the likelihood of occurrence of very large errors by bounding their variance (mean square).
In other words, Hypotheses \eqref{eq:zeromean} and \eqref{eq:MSE} simply mean that the players' individual gradient estimates $\hat\payv_{\play}$ are \emph{unbiased and bounded in mean square}, i.e.
\begin{subequations}
\label{eq:estimates}
\begin{flalign}
\label{eq:unbiased}
&\exof[\big]{\hat\payv_{\play}(n) \given \bX(n)}
	= \payv_{\play}(\bX(n)),
	\\
\label{eq:vbound}
&\exof[\big]{ \dnorm{\hat\payv_{\play}(n)}^{2} \given \bX(n) }
	\leq \vbound_{\play}^{2}
%	< +\infty
	\quad
	\text{for some $\vbound_{\play}>0$}.
\end{flalign}
\end{subequations}

Even though \eqref{eq:MSE} will be our main error control assumption, it will also be useful to consider the more refined hypothesis:
\begin{enumerate}
[(H2$'$)]
\setcounter{enumi}{2}
\item
\label{hyp:subexp}
\emph{Subexponential moment growth:}
for all $p\in\N$,
\begin{equation}
\tag{H2$'$}
\label{eq:subexp}
\exof[\big]{\dnorm{\bZ(n)}^{p} \given \bX(n)}
	\leq \frac{p!}{2} \noisedev^{p}
%\dnorm{\bZ(n)}
%	\leq \noisedev_{n}
	\quad
	\text{for some $\noisedev>0$}.
\end{equation}
\end{enumerate}
Clearly, \eqref{eq:MSE} is implied by \eqref{eq:subexp} but the latter may fail in certain heavy-tailed error distributions (such as Pareto-tailed ones).
From a practical point of view, this is not particularly restrictive as \eqref{eq:subexp} continues to apply to a wide range of practical scenarios (including all Gaussian and sub-Gaussian error distributions), and is considerably more general than the ``finite errors'' assumptions studied in the literature \cite{SS11}.
%given that real-world feedback and measurements are often \emph{de facto} bounded, Hypothesis \eqref{eq:subexp} is not particularly restrictive and continues to apply to a wide range of practical scenarios and systems (including all Gaussian and sub-Gaussian distributions).

%----------------------------------------------------------------------
%%% ANALYSIS
%----------------------------------------------------------------------
\section{Convergence Analysis and Implementation}
\label{sec:analysis}
%----------------------------------------------------------------------
%%% ANALYSIS
%----------------------------------------------------------------------
% !TEX root = ./Main.tex

By construction, the recursion \eqref{eq:MXL} with gradient feedback satisfying Hypotheses \eqref{eq:zeromean} and \eqref{eq:MSE} enjoys the following desirable properties:
\begin{enumerate}
[\noindent (P1)]
\item
\emph{Distributedness:}
players only require individual gradient information as feedback.
\item
\emph{Robustness:}
the players' feedback could be stochastic, imperfect, or otherwise perturbed by random noise.
\item
\emph{Statelessness:}
players do not need to know the state of the system.
\item
\emph{Reinforcement:}
players tend to increase their individual utilities.
\end{enumerate}
The above shows that \eqref{eq:MXL} is a promising candidate for learning in games and distributed optimization problems of the general form \eqref{eq:game}.
Accordingly, our aim in this section will be to examine the long-term convergence properties of \eqref{eq:MXL} and its practical implementation attributes.

\subsection{Convergence}
We begin by showing that the algorithm's termination states are \aclp{NE}:

\begin{theorem}
\label{thm:stationary}
Assume that Algorithm \ref{alg:MXL} is run with
a decreasing step-size sequence $\step_{n}$ such that $\sum_{n=1}^{\infty} \step_{n}^{2} < \sum_{n=1}^{\infty} \step_{n} = \infty$
and
gradient observations satisfying \eqref{eq:zeromean} and \eqref{eq:MSE}.
If $\bX(n)$ converges, it does so to a \acl{NE} of the game \as.
\end{theorem}

%\begin{IEEEproof}
%See Appendix \ref{app:stationary}.
%\end{IEEEproof}

The proof of Theorem \ref{thm:stationary} is presented in detail in Appendix \ref{app:stationary} and is essentially by contradiction.
To provide some intuition, if the limit of \eqref{eq:MXL} is not a \acl{NE}, at least one player of the game must be dissatisfied,  thus experiencing a repelling drift \textendash\ in the limit and on average.
Owing to the algorithm's exponentiation step, this ``repulsion'' can be quantified via the so-called \emph{von Neumann} (or \emph{quantum}) \emph{entropy} \cite{Ved02};
then, by using the law of large numbers, it is possible to control the impact of the noise and show that this entropy diverges to infinity, thus obtaining the desired contradiction.

\smallskip
Stated concisely, Theorem \ref{thm:stationary} shows that if \eqref{eq:MXL} converges, it converges to a \acl{NE}.
However, the theorem does not provide any guarantee that the algorithm converges in the first place. A sufficient condition for convergence based on equilibrium stability is given below:

\begin{theorem}
\label{thm:global}
Assume that Algorithm \ref{alg:MXL} is run with
a step-size sequence $\step_{n}$ such that $\sum_{n=1}^{\infty} \step_{n}^{2} < \sum_{n=1}^{\infty} \step_{n} = \infty$
and
gradient observations satisfying \eqref{eq:zeromean} and \eqref{eq:MSE}.
If $\eq$ is globally stable, then $\bX(n)$ converges to $\eq$ \as.
\end{theorem}

The proof of Theorem \ref{thm:global} is given in Appendix \ref{app:global}.
In short, it comprises the following steps:
First, we consider a deterministic, continuous-time variant of \eqref{eq:MXL} and we show that globally stable equilibria are global attractors of said system.
To do this, we introduce a matrix version of the so-called \emph{Fenchel coupling} \cite{MS16} and we show that it plays the role of a Lyapunov function in continuous time.
Subsequently, we derive the evolution of the discrete-time stochastic system \eqref{eq:MXL} by using the method of stochastic approximation \cite{Ben99} and the theory of concentration inequalities \cite{dlP99} to control the aggregate error between continuous and discrete time.

\smallskip

From a practical point of view, an immediate corollary of Theorem \ref{thm:global} is the following second-derivative convergence test:
%(similar to the second-derivative test in ordinary calculus):

\begin{corollary}
\label{cor:global}
If the game's Hessian matrix $\hessmat(\bX)$ is negative-definite for all $\bX\in\feas$, Algorithm \ref{alg:MXL} converges to the game's \textup(necessarily\textup) unique \acl{NE} \as.
\end{corollary}

The above results show that \eqref{eq:MXL} converges to stable equilibria under very mild assumptions on the underlying stochasticity (zero-mean errors and finite conditional variance).
The following theorem shows that, under the slightly sharper assumption \eqref{eq:subexp}, local stability implies local convergence with arbitrarily high probability:

\begin{theorem}
\label{thm:local}
Assume that Algorithm \ref{alg:MXL} is run with
gradient observations satisfying \eqref{eq:zeromean} and \eqref{eq:subexp},
and small enough step-sizes $\step_{n}$ such that $\sum_{n=1}^{\infty} \step_{n}^{1+q} < \sum_{n=1}^{\infty} \step_{n} = \infty$ for some $q\in(0,1)$.
If $\eq$ is \textup(locally\textup) stable, then it is locally attracting with arbitrarily high probability;
specifically, for every $\eps>0$, there exists a neighborhood $U_{\eps}$ of $\eq$ such that
\begin{equation}
\label{eq:prob-local}
\probof{\bX(n) \to \eq \given \bX(0) \in U_{\eps}}
	\geq 1-\eps.
\end{equation}
%whenever $\bX(0)\in U_{\eps}$.
\end{theorem}

Theorem \ref{thm:local} is proven in Appendix \ref{app:local}, building on the (global) proof of Theorem \ref{thm:global}.
The key difference with Theorem \ref{thm:global} is that, since we only assume the existence of a \emph{locally stable} equilibrium $\eq$, the drift of \eqref{eq:MXL} has no reason to point towards $\eq$ globally.
As a result, $\bX(n)$ may exit the basin of $\eq$ in the presence of high uncertainty.
However, by invoking \eqref{eq:subexp} and the Borel\textendash Cantelli lemma, it is possible to show that this happens with arbitrarily small probability, leading to the probabilistic convergence result \eqref{eq:prob-local}.

As in the case of globally stable equilibria, Theorem \ref{thm:local} leads to the following easy-to-check condition for local convergence:
\begin{corollary}
\label{cor:local}
Let $\eq$ be a \acl{NE} of $\game$ such that $\hessmat(\eq) \prec 0$.
Then \eqref{eq:MXL} converges locally to $\eq$ with arbitrarily high probability.
\end{corollary}

%\VB{Corollary 2 is true given the assumptions of Theorem 4 (regarding gamma for ex)? and Corrolary 1 given the assumptions in Theorem 3? Should this be specified maybe for clarity? }
%\PM{Meeeh\dots}

\subsection{Rate of convergence}

Combining Theorems \ref{thm:stationary}\textendash \ref{thm:local} gives a fairly complete picture of the \emph{qualitative} convergence properties of the exponential learning algorithm \eqref{eq:MXL}:
the only possible end-states of \eqref{eq:MXL} are \aclp{NE}, and the existence of a globally (resp.~locally) stable \acl{NE} implies the algorithm's global (resp.~local) convergence to said equilibrium. On the other hand, these theorems do not address the \emph{quantitative} aspects of the algorithm's long-term behavior, such as its rate of convergence.
In what follows, we study precisely this question.

We begin by introducing the so-called \emph{quantum Kullback\textendash Leibler divergence} (or \emph{von Neumann relative entropy}) to measure distances on $\feas$ \cite{Ved02}.
Specifically, the quantum divergence between $\bX$ and $\eq$ is defined as
\begin{equation}
\label{eq:Bregman}
\breg(\eq,\bX)
	= \trof{\eq (\log\eq - \log\bX)}.
\end{equation}
By Klein's inequality \cite{Ved02}, $\breg(\eq,\bX) \geq 0$ with equality if and only if $\bX=\eq$, so $\dkl(\eq,\bX)$ represents a (convex) measure of the distance between $\bX$ and $\eq$.
With this in mind, we introduce below the following quantitative measure of stability:
\begin{definition}
\label{def:stable-strong}
Given $B>0$, $\eq$ is called \emph{$B$-strongly stable} if
\begin{equation}
\label{eq:stable-strong}
\trof{(\bX - \eq)\,\payv(\bX)}
	\leq -B\,\breg(\eq,\bX)
	\quad
	\text{for all $\bX\in\feas$}.
\end{equation}
\end{definition}

Under this refinement of equilibrium stability,%
\footnote{Since $\breg(\eq,\bX) \geq 0$ with equality if and only if $\bX = \eq$, it follows that strongly stable states are automatically stable.}
we obtain the following quantitative result:

\begin{theorem}
\label{thm:convrate}
Assume that Algorithm \ref{alg:MXL} is run with the step-size sequence $\step_{n} = \step/n$
and
gradient observations satisfying \eqref{eq:zeromean} and \eqref{eq:MSE}.
If $\eq$ is $B$-strongly stable and $\step > B^{-1}$, we have
\begin{equation}
\label{eq:convrate}
\exof{\breg(\eq,\bX(n))}
	\leq \frac{\step^{2} \vbound^{2}}{B\step - 1} \frac{1}{n}.
\end{equation}
\begin{subequations}
In particular, if $\eq$ lies in the interior of $\feas$, then
\begin{equation}
\label{eq:convrate-int}
\exof{\norm{\bX(n) - \eq}}
	= \bigoh(n^{-1/2}).
\end{equation}
Otherwise, if $\eq$ is an extreme point of $\feas$, we have
\begin{equation}
\label{eq:convrate-ext}
\exof{\norm{\bX(n) - \eq}}
	= \bigoh(n^{-1}).
\end{equation}
\end{subequations}
\end{theorem}

The explicit bounds of Theorem \ref{thm:convrate} (proven in Appendix \ref{app:rates}) have several interesting consequences:

\begin{enumerate}
[\itshape i\upshape)]
\item
The rate of convergence to interior versus extreme states of $\feas$ is substantially different because the von Neumann divergence grows quadratically when $\eq$ is interior and linearly when $\eq$ is extreme.
Intuitively, the reason for this is that, in the case of interior states, the algorithm has to slow down considerably near $\eq$ in order for random oscillations around $\eq$ to dissipate.
On the other hand, in the case of extreme points, the algorithm has a constant drift towards the boundary of $\feas$ and there are no possibilities of overshooting (precisely because the equilibrium state in question is an extreme point of $\feas$).
As a result, random oscillations become irrelevant near extreme points, thus allowing for faster convergence in that case. 

\smallskip

\item
Thanks to the explicit expression \eqref{eq:convrate}, we see that if the stability constant $B$ can be estimated ahead of time, we can attain the convergence rate
\begin{equation}
\label{eq:convrate-opt}
\exof{\breg(\eq,\bX(n))}
	\leq \frac{4\vbound^{2}}{B^{2}n},
\end{equation}
achieved for the optimized step-size sequence $\step_{n} = 2/(Bn)$.
%Explicit bounds of this form can also be derived for step-size sequences of the form $\step_{n} \propto n^{-\alpha}$ for $\alpha<1$, but the $\bigoh(1/n)$ convergence rate of \eqref{eq:convrate} is the fastest possible, so we do not present them.
It is also possible to estimate the algorithm's convergence rate for equilibrium states that are only locally strongly stable.
However, given that the algorithm's convergence is probabilistic in that case, the resulting bounds are also probabilistic and, hence, more complicated to present.
%instead of deriving them here, we examine this question via numerical simulations in Section \ref{sec:numerics}.
\end{enumerate}

\subsection{Practical implementation}
\label{par:implementation}

We close this section with a discussion of certain issues pertaining to the practical implementation of Algorithm \ref{alg:MXL}:

\paragraph{Updates and feasibility}
\label{par:feasible}

As we discussed earlier, the exponentiation/normalization step of Algorithm \ref{alg:MXL} ensures that the players' action variables $\bX_{\play}(n)$ satisfy the game's feasibility constraints \eqref{eq:feasible} at each stage $n$.
In the noiseless case ($\noise=0$), feasibility is automatic because, by construction, $\payv_{\play}$ (and, hence, $\bY_{\play}$) is Hermitian.
%and respects the block structure of $\feas_{\play}$ (namely $\exp(\bY_{\play})$ is positive-definite and block-similar to the fixed model matrix $\bB_{\play}$).
In the presence of noise however, there is no reason to assume that the estimates $\hat\payv_{\play}$ are Hermitian,
%or block-diagonal,
so the updates $\bX_{\play}$ may also fail to be feasible.
To rectify this, we tacitly assume that
each player corrects such errors by replacing $\hat\payv_{\play}$ with $(\hat\payv_{\play} + \hat\payv_{\play}^{\dag})/2$.
Since this error-correcting operation is linear in its input, Hypotheses \eqref{eq:zeromean} and \eqref{eq:MSE} continue to apply, so our analysis and proofs hold as stated.
%the feedback to \eqref{eq:MXL} is pre-processed by each player to correct such errors as follows:
%\begin{inparaenum}
%[\itshape a\upshape)]
%\item
%$\hat\payv_{\play}$ is replaced by $(\hat\payv_{\play} + \hat\payv_{\play}^{\dag})/2$;
%and
%\item
%any elements of the resulting matrix that do not conform to the block-diagonal structure of $\feas_{\play}$ are chopped off.
%\end{inparaenum}
%Since these error-correcting operations are both linear in their input, Hypotheses \eqref{eq:zeromean} and \eqref{eq:MSE} continue to apply to the post-processed feedback, so our analysis and proofs hold as stated.

\paragraph{On the step-size sequence $\step_{n}$}
\label{par:stepsize}

Using a decreasing step-size $\step_{n}$ in \eqref{eq:MXL} may appear counter-intuitive because it implies that new information enters the algorithm with decreasing weights.
As evidenced by Theorem \ref{thm:convrate}, the reason for this is that
%when $\bX(n)$ approaches an equilibrium state,
a constant step-size might cause the process to overshoot and lead to oscillatory behavior around the algorithm's end-state.
In the deterministic regime, these oscillations can be dampened by using forward-backward splitting or accelerated descent methods \cite{Nes04}.
However, in the presence of noise, the use of a decreasing step-size is essential in order to dissipate measurement noise and other stochastic effects, explaining why it is not possible to improve on \eqref{eq:convrate} by using a constant step.

\paragraph{Computational complexity and numerical stability}
\label{par:complexity}

From the point of view of computational complexity, the bottleneck of each iteration of Algorithm \ref{alg:MXL} is the matrix exponentiation step $\bY_{\play} \mapsto \exp(\bY_{\play})$.
%\footnote{For the moment, we are assuming that gradients are calculated at a relatively small cost.
%In an abstract setting, this is justified because such estimates are obtained via an oracle call;
%otherwise, the validity of this assumption in practical scenarios is discussed in detail in the case studies that we examine in Part II of our paper.}
Since matrix exponentiation has the same complexity as matrix multiplication, this step has polynomial complexity with a low degree on the input dimension of $\bX_{\play}$ \cite{MvL03}.
In particular, each exponentiation requires $\bigoh(\tx_{\play}^{\omega})$ floating point operations, where the complexity exponent can be as low as $\omega = 2.373$ if the players  employ fast Coppersmith\textendash Winograd matrix multiplication methods \cite{DS13}. 
%Moreover, if the model matrix $\bB_{\play}$ is not composed of a single block, this cost can be further reduced to $\bigoh(SB_{\play}^{\omega})$ where $S$ is the number of blocks and $B_{\play}$ is the maximum block size;
%in particular, if $\bB_{\play}$ is diagonal (i.e. $\feas_{\play}$ is a simplex), the operation is linear in the dimensionality of the input.
%\VB{I'm not sure what exactly is the model matrix $\bB_{\play}$... do we need such details (involving new notations)? or can we just  simply say smth like "Moreover, the special structure of the matrices (e.g., block or diagonal matrices) can be exploited to further reduce the computational cost." and eventually add the refs? Also, out of curiosity: do we actually use the Coppersmith\textendash Winograd thing? Is it a practical implemented method?}
%\PM{Removed the model matrix stuff.
%I have no idea what Mathematica/Matlab are using. CW is not easily implementable, but there are implementable methods with complexity around 2.45 I think?
%A colleague explained all that to me once, but I don't remember now\dots}

Finally, regarding the numerical stability of Algorithm \ref{alg:MXL}, the only possible source of arithmetic errors is the exponentiation/normalization step.
Indeed, if the eigenvalues of $\bY_{\play}$ are large, this step could incur an overflow where both the numerator and the denominator evaluate to machine infinity.
This potential instability can be fixed as follows:
If $y_{\play}$ denotes the largest eigenvalue of $\bY_{\play}$ and we let $\bY_{\play}' = \bY_{\play} - y_{\play} \bI$, the algorithm's exponentiation step can be rewritten as:
\begin{equation}
\label{eq:exp-stable}
\bX_{\play}
	\leftarrow \frac{\exp(\bY_{\play}')}{\exp(-y_{\play}) + \norm{\exp(\bY_{\play}')}}.
\end{equation}
Thanks to this shift, the elements of the numerator are now bounded from above by $1$ (because the largest diagonal element of $\bY_{\play}'$ is $e^{y_{\play} - y_{\play}} = 1$), so there is no danger of encountering a numerical indeterminacy of the type $\mathtt{Inf}/\mathtt{Inf}$.
Thus, to avoid computer arithmetic issues, we employ the stable expression \eqref{eq:exp-stable} in all numerical implementations of Algorithm \ref{alg:MXL}.

\paragraph{Asynchronous implementation}
\label{par:asynchronicity}

{
An implicit assumption in \eqref{eq:MXL} is that user updates \textendash\ albeit local \textendash\ are concurrent.
This can often be achieved via a global update timer that synchronizes the players' updates;
however, in a fully decentralized setting, even this degree of coordination may be challenging to attain.
Thus, to overcome this synchronicity limitation, we discuss below an asynchronous variant of Algorithm \ref{alg:MXL} where each player updates his action based on an individual, independent schedule.}

Specifically, assume that each player $\play\in\playerset$ has an individual timer that triggers an update event, i.e. a request for gradient feedback and, subsequently, an update of $\bX_{\play}$.
%\footnote{More precisely, we assume here that $\tau_{k}\from\N\to\R_{+}$ is an increasing (and possibly random) sequence such that $\tau_{k}(n)$ marks the instance in time at which the $k$-th user updates his covariance matrix $\bQ_{k}$ for the $n$-th time \textendash\ so $\bQ_{k}$ changes at $\tau_{k}(n)$ and remains constant throughout $[\tau_{k}(n),\tau_{k}(n+1))$.}
%Thus,
%at every tick of $\tau_{\play}$, player $\play$ gets an estimate of $\payv_{\play}$ and updates $\bX_{\play}$.
Of course, the players' gradient estimates $\hat \payv_{\play}$ will then be subject to delays and asynchronicities (in addition to noise), so the update structure of Algorithm \ref{alg:MXL} must be modified appropriately.
To that end, let $\playset_{n} \subseteq \playset$ be the set of players that update their actions at the $n$-th overall update event (typically $\abs{\playset_{n}}=1$ if players update at random times),
and let $d_{\play}(n)$ be the corresponding number of epochs that have elapsed since the last update of player $\play$.
%and let $\tau(k)$ denote the time at which this \update occurs.
%(so $\tau(n)$ is an ``aggregate'' counter for the transmitters' \updates).
%Furthermore, let $n_{k}$ denote the update counter of the $k$-th user at the $n$-th overall \update and let $n_{0}$ denote the value of the corresponding \feedback counter of the receiver.%
%\footnote{In formal mathematical language, $n_{j} = \sup\{m : \tau_{j}(m) \leq \tau(n) \}$, $j = 0,1,\dotsc,K$, and $d_{k}(n) = n - n(\tau_{k}(n_{k}(\tau_{0}(n_{0}))))$.}
%Dually, given some $t\geq0$, let $n^{\sharp}(t) = \sup\{n : \tau(n) \leq t\}$ denote the total number of \updates that have occurred up to time $t$, and define the corresponding counting functions $n_{0}^{\sharp}(t)$ and $n_{k}^{\sharp}(t)$ for \feedbacks at the receiver and \updates for user $k$.
%Finally, let the delay variable $d_{k}(n)$ denote the number of \updates that have elapsed between the $n$-th \update and the last \update of user $k$ before the most recent \feedback.%
%\footnote{More formally (but far less intuitively), $d_{k}(n) = n - n^{\sharp}(\tau_{k}(n_{k}^{\sharp}(\tau_{0}(n_{0}^{\sharp}(\tau(n))))))$.}
We then obtain the following asynchronous variant of \eqref{eq:MXL}:
\begin{equation}
\label{eq:MXL-async}
\begin{aligned}
\bY_{\play}(n+1)
	&= \bY_{\play}(n) + \step_{n_{\play}} \one\braces{\play\in\playset_{n}} \cdot \hat\payv_{\play}(n),
	\\
\bX_{\play}(n+1)
	&= A_{\play} \frac{\exp(\bY_{\play}(n+1))}{1 + \norm{\exp(\bY_{\play}(n+1))}},
\end{aligned}
\end{equation}
where
%$n_{\play} = \sum_{j=1}^{n} \one\braces{\play\in\playerset_{j}}$
$n_{\play}$ denotes the number of updates performed by player $\play$ up to epoch $n$ while the (asynchronous) estimate $\hat\payv_{\play}(n)$ satisfies
\begin{equation}
\label{eq:grad-multi}
%\tag{\ref*{eq:zeromean}$'$}
\exof{\hat\payv_{\play}(n)}
	= \payv_{\play}(\bX_{1}(n - d_{1}(n)),\dotsc,\bX_{\players}(n - d_{\players}(n))).
\end{equation}
Assuming that the delays $d_{\play}(n)$ are bounded and the players' update rates are strictly positive (in the sense that $\liminf_{n\to\infty} n_{\play}/n > 0$), it is possible to employ the analysis of \cite{CGM15} to show that Theorems \ref{thm:stationary}\textendash\ref{thm:local} remain true under the asynchronous variant \eqref{eq:MXL-async}.
However, a detailed analysis would take us too far afield, so we relegate it to future work.

%----------------------------------------------------------------------
%%% NUMERICS
%----------------------------------------------------------------------
\section{Numerical Results}
\label{sec:numerics}
%----------------------------------------------------------------------
%%% NUMERICS
%----------------------------------------------------------------------
% !TEX root = ./Main.tex

\begin{table}[t]
\caption{wireless network simulation parameters}
\label{tab:parameters}
\renewcommand{\arraystretch}{1.1}
\centering
\footnotesize
%----------------------------------------------------------------------
%%% PARAMETERS
%----------------------------------------------------------------------
% !TEX root = ./Main.tex

\begin{tabular}{|c|c|}
\hline
\textbf{Parameter}
	&\textbf{Value}
%	&\textbf{Parameter}
%	&\textbf{Value}
	\\
	\hline
Cell size (rectangular)
	&$1\,\km$
\\
\hline
User density
	&$500\,\textrm{users}/\km^{2}$
\\
\hline
Time frame duration
	&$5\,\ms$
\\
\hline
Wireless propagation model
	&COST Hata
%	&BS/MS antenna height
%	&$32\,\meter$ / $1.5\,\meter$
\\
\hline
Central frequency
	&$2.5\,\ghz$
\\
\hline
Total bandwidth
	&$11.2\,\mhz$
\\
\hline
\acs{OFDM} subcarriers
	&$1024$
\\
\hline
Subcarrier spacing
	&$11\,\khz$
\\
\hline
Spectral noise density ($20\,^{\circ}\textrm{C}$)
	&$-174\,\dbm/\hz$
%\\
%\hline
%User speed
%	&$[3,130]\,\kmh$
\\
\hline
Maximum transmit power
	&$\pmax = 33\,\dbm$
\\
\hline
Non-radiative power
	&$\pc = 20\,\dbm$
\\
\hline
Transmit antennas per device
	&$\tx=4$
\\
\hline
Receive antennas per link
	&$\rx=8$
\\
\hline
\end{tabular}
\end{table}

\begin{figure*}[htbp]
\footnotesize
\subfigure{\label{fig:static-perfect}%
\includegraphics[width=.48\textwidth]{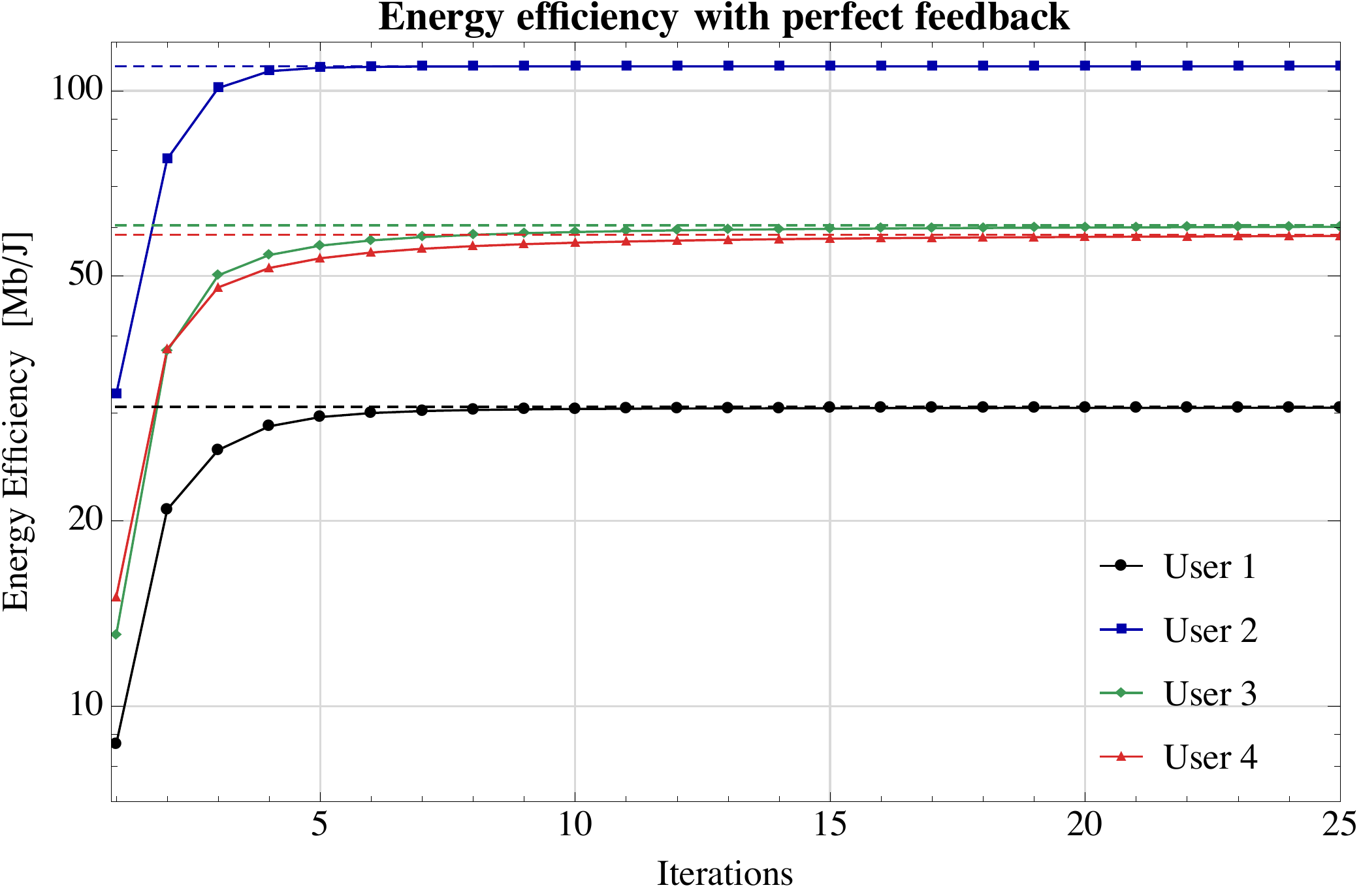}}
\hfill
\subfigure{\label{fig:static-z25}%
\includegraphics[width=.48\textwidth]{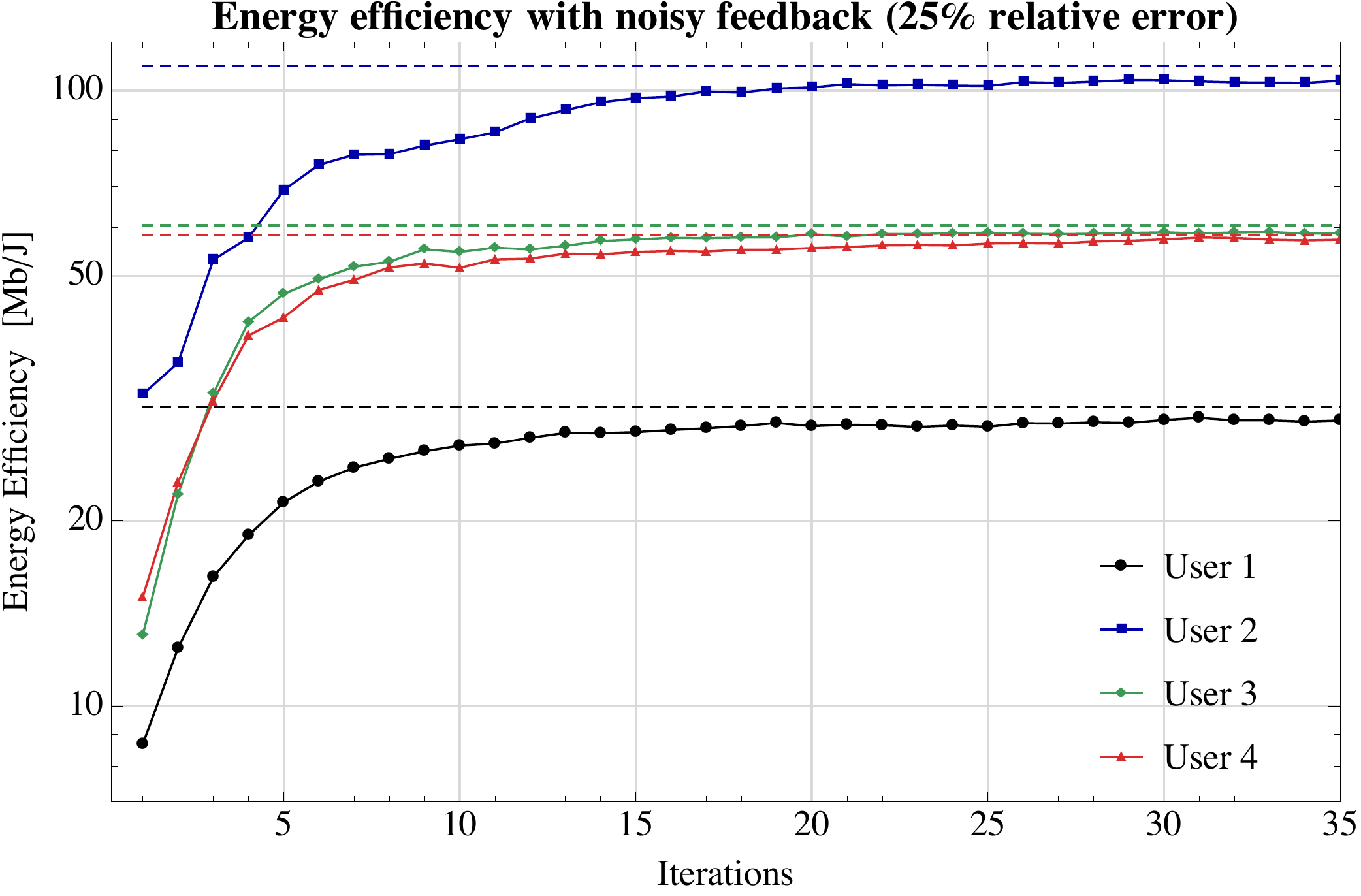}}\:
\\
\subfigure{\label{fig:static-z50}%
\includegraphics[width=.48\textwidth]{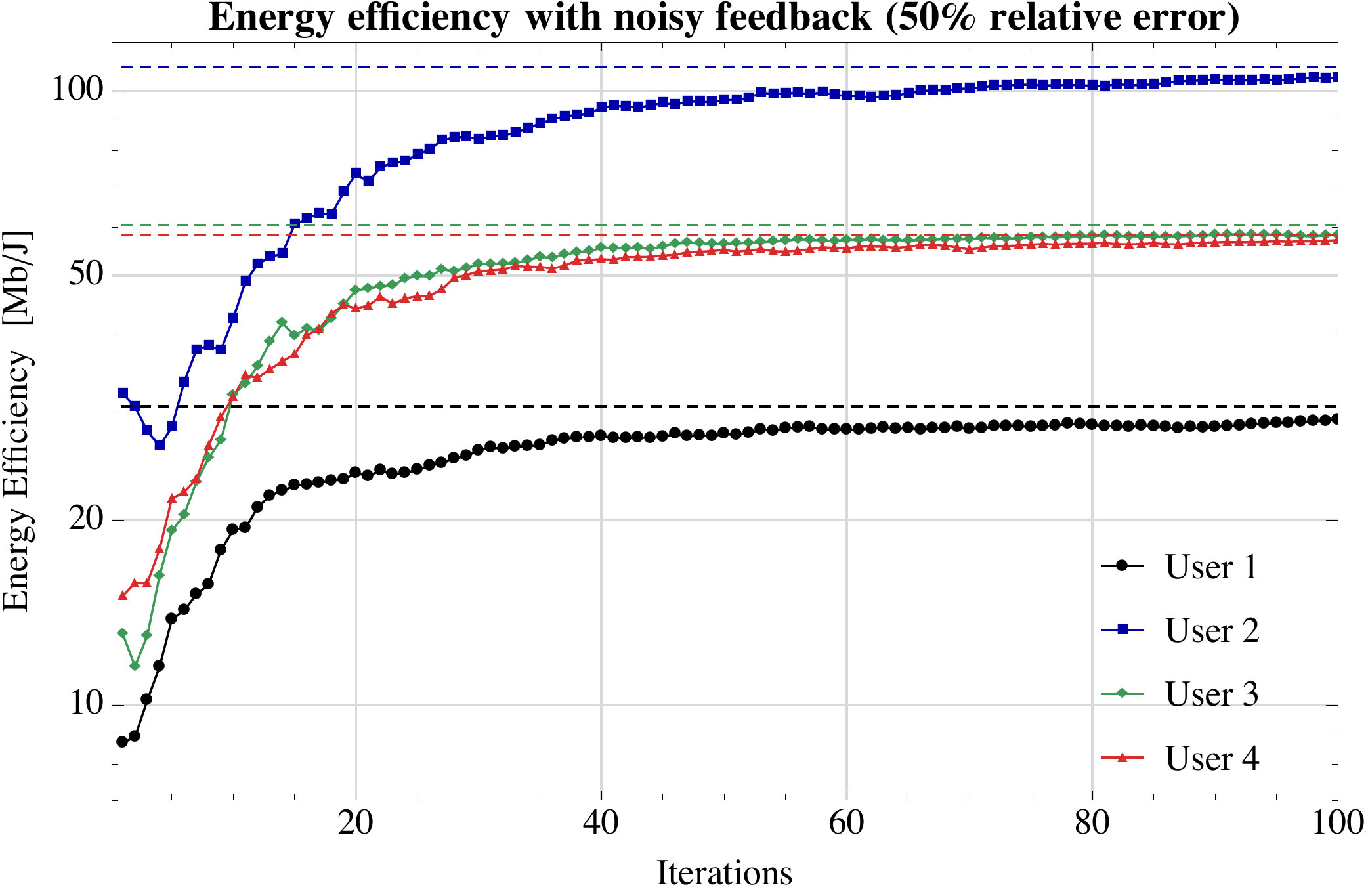}}
\hfill
\subfigure{\label{fig:static-z100}%
\includegraphics[width=.48\textwidth]{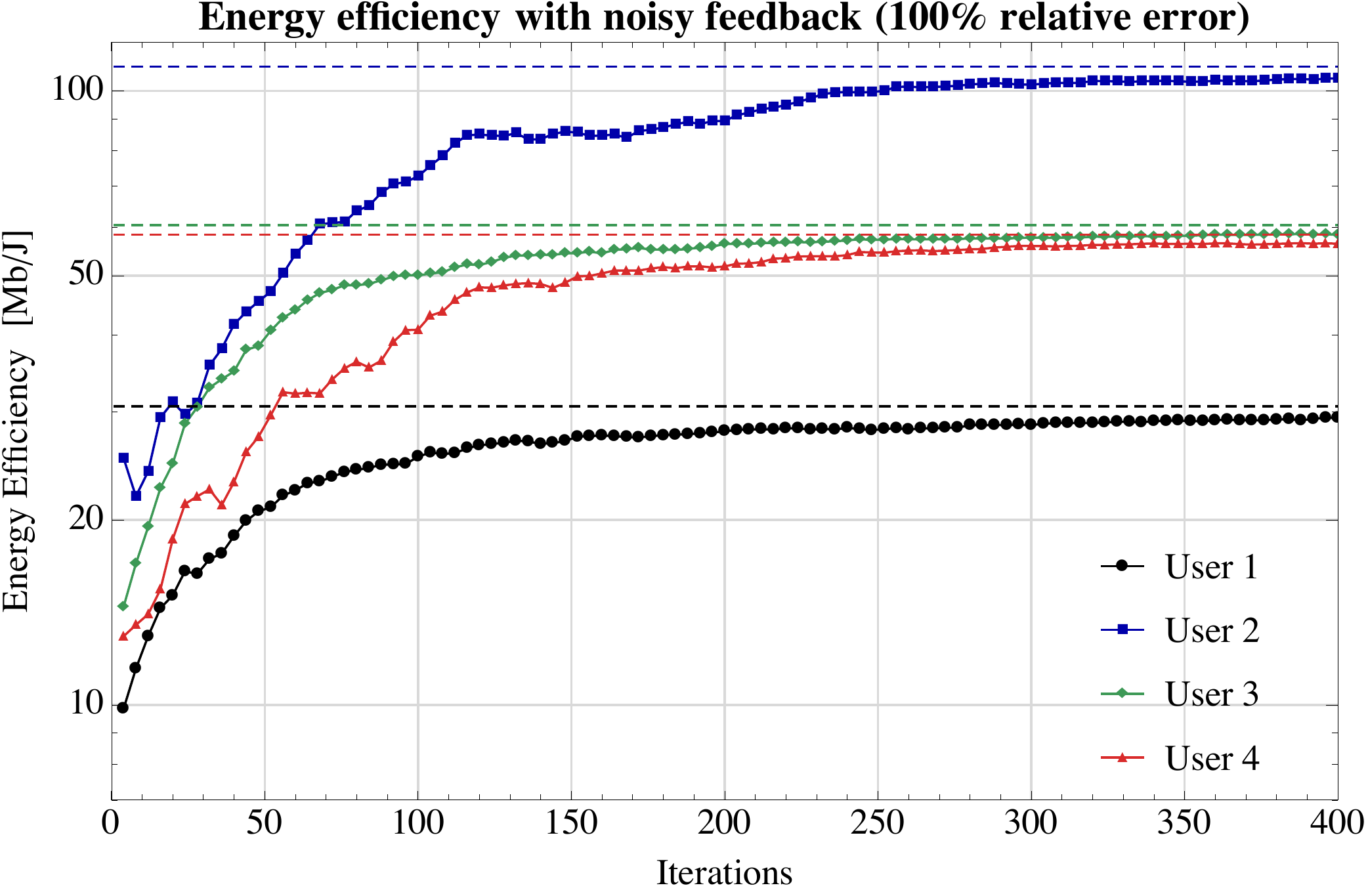}}%
\caption{Performance of \ac{MXL} in the presence of noise.
In all figures, we plot the transmit \acl{EE} of wireless users that employ Algorithm \ref{alg:MXL} in a wireless network with parameters as described in the main text (to reduce graphical clutter, we only plotted $4$ users with diverse channel characteristics).
In the absence of noise (upper left), the system converges to a stable \acl{NE} state (unmarked dashed lines) within a few iterations.
The convergence speed of \ac{MXL} is slower in the presence of noise but the algorithm remains convergent under very high degrees of uncertainty (up to relative error levels of $100\%$; bottom right).}
\vspace{-2ex}
\label{fig:static}
\end{figure*}

\begin{figure*}[t]
\footnotesize
\subfigure[Channel gain evolution for different user velocities]{\label{fig:varying-channels}%
\includegraphics[width=.48\textwidth]{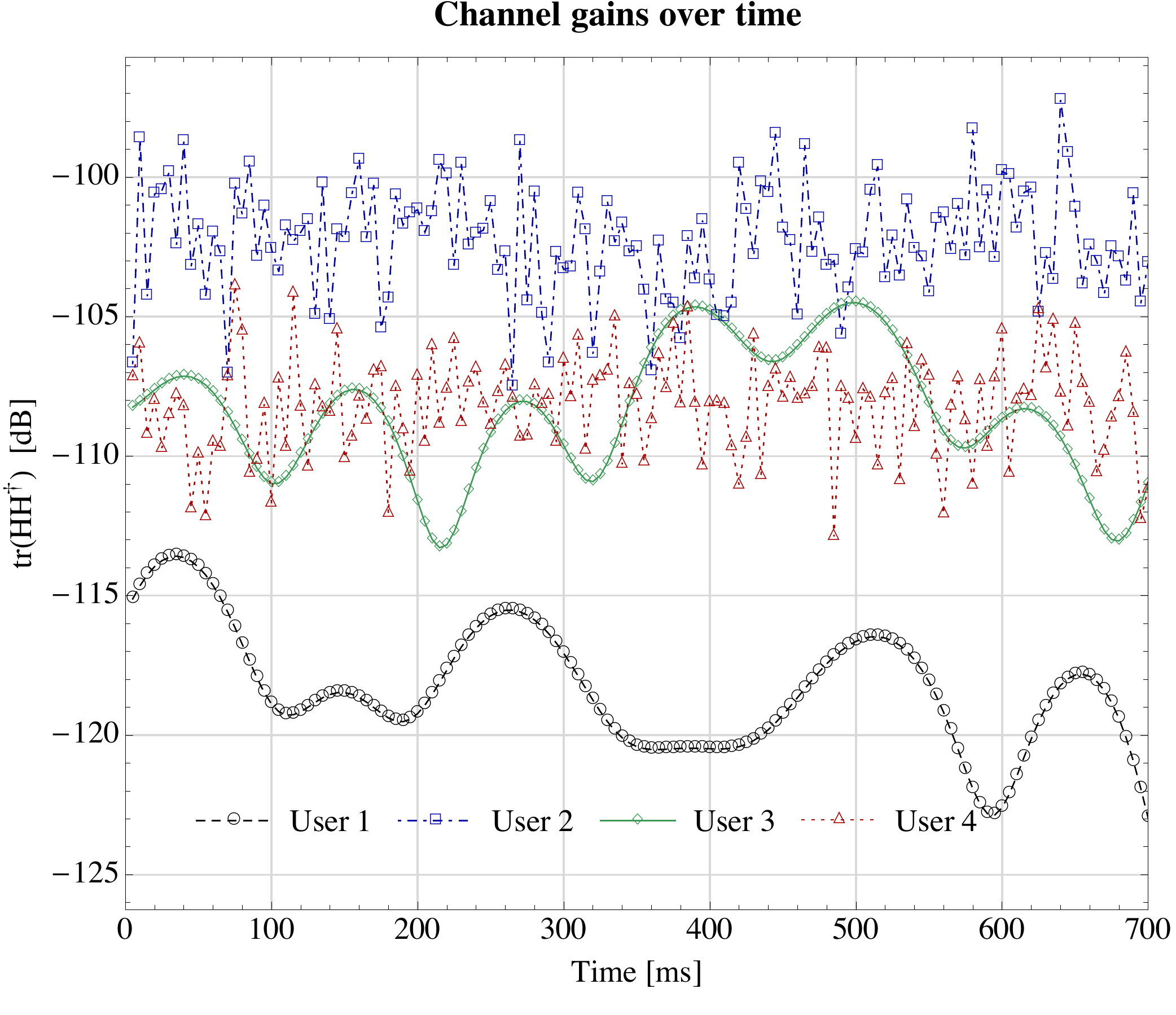}}
\hfill
\subfigure[Equilibrium tracking under mobility]{\label{fig:varying-tracking}%
\includegraphics[width=.48\textwidth]{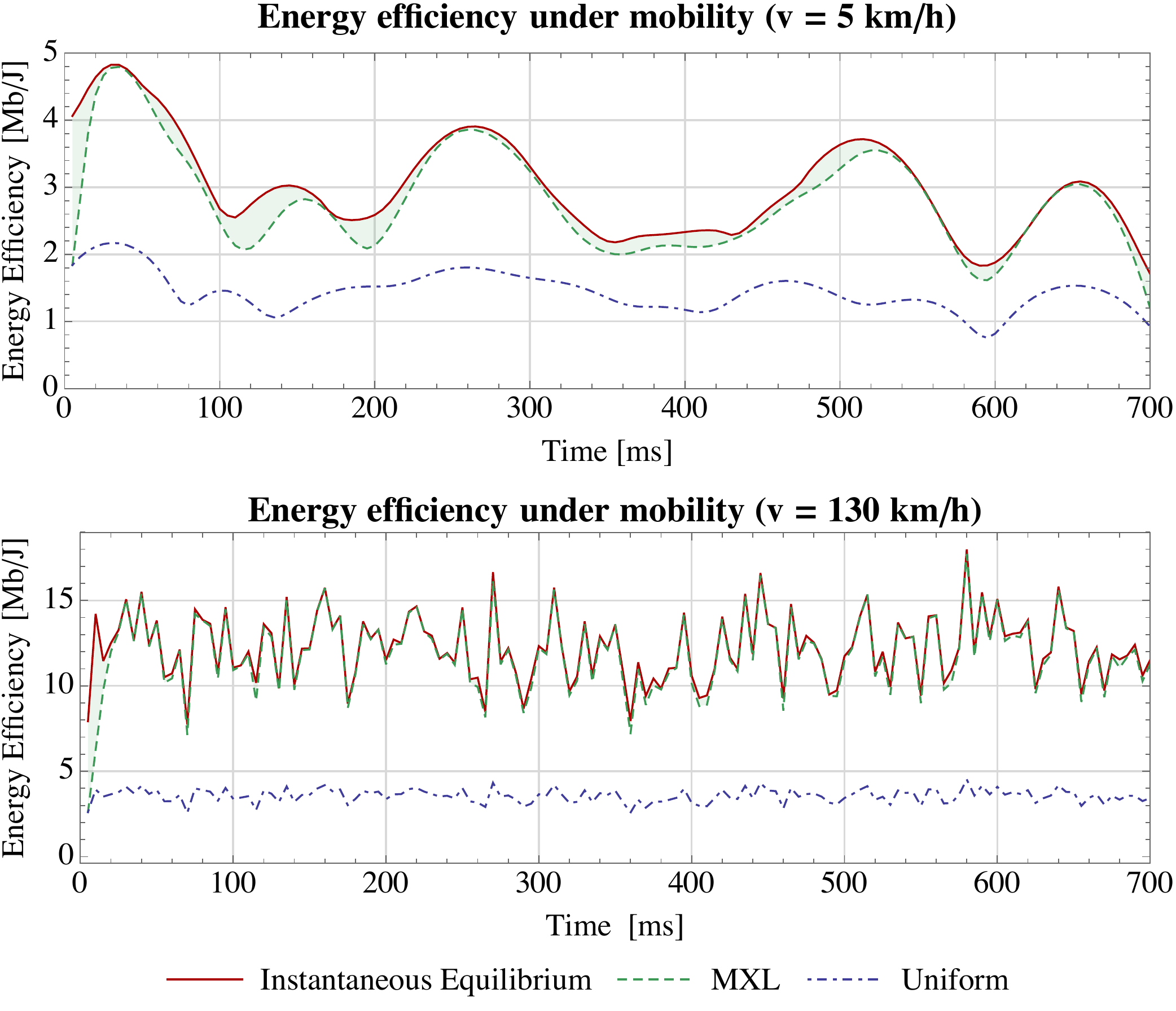}}
\caption{Performance of \ac{MXL} in a dynamic network environment with mobile users moving at $v \in \{3, 5, 30,130\}\,\kmh$.
%Despite the network's fully dynamic character,
%the \ac{MXL} algorithm achieves a no-regret state within a few transmission frames.
%Furthermore, as we see in Figs.~\ref{fig:varying-EE-slow} and \ref{fig:varying-EE-fast},
The users' achieved \acl{EE} tracks the system's (evolving) equilibrium remarkably well, even under rapidly changing channel conditions.}
\label{fig:varying}
\end{figure*}

In this section, we assess the performance of the \ac{MXL} algorithm via numerical simulations.
For concreteness, we focus on the case of transmit \acl{EE} maximization in practical multi-user \ac{MIMO} networks (cf. Section \ref{sec:EE}), but our conclusions apply to a wide range of parameters and scenarios.
%{\color{blue} As we have already argued, this problem has not yet been tackled in the literature until now.}
To the best of our knowledge, this comprises the first distributed solution scheme for general multi-user/multi-antenna/multi-carrier networks under imperfect feedback/\ac{CSI} and mobility considerations.

Our basic network setup consists of a macro-cellular \acs{OFDMA} wireless network with access points deployed on a rectangular grid with cell size $1\,\km$ (for a quick overview of simulation parameters, see Table \ref{tab:parameters}).
Signal transmission and reception occurs over a $10\,\mhz$ band divided into $1024$ subcarriers around a central frequency of $f_{c} = 2.5\,\ghz$.
We further assume a frame-based \ac{TDD} scheme with frame duration $T_{f} = 5\,\ms$:
transmission takes place during the \acl{UL} phase while the network's access points process the received signal and provide feedback during the \acl{DL} phase.
Finally, signal propagation is modeled after the widely used COST Hata model \cite{Hat80,COST99} with spectral noise density equal to $-174\,\dbm/\hz$ at $20\,^{\circ}\textrm{C}$.

The network is populated by wireless transmitters (users) following a homogeneous Poisson point process with intensity $\rho = 500\,\textrm{users}/\km^{2}$.
Each wireless transmitter is further assumed to have $\tx=4$ transmit antennas, a maximum transmit power of $\pmax = 40\,\dbm$ and circuit (non-radiative) power consumption of $\pc = 20\,\dbm$.
In each cell, \ac{OFDM} subcarriers are allocated to wireless users randomly so that different users are assigned to disjoint carrier sets.
We then focus on a set of $\players=25$ users, each located at a different cell of a $5\times5$ square cell cluster and sharing $\carriers=8$ common subcarriers.
Finally, at the receiver end, we consider $\rx=8$ receive antennas per connection and a receiver noise figure of $7\,\db$.

To assess the performance and robustness of the \ac{MXL} algorithm, we first focus on a scenario with stationary users and static channel conditions.
Specifically, in Fig.~\ref{fig:static}, each user runs Algorithm \ref{alg:MXL} with a variable step-size $\step_{n} \sim n^{-1/2}$ and initial transmit power $P_{0} = \pmax/2 = 26\,\dbm$ (allocated uniformly across different antennas and subcarriers),
and we plot the users' transmit \acl{EE} over time.
For benchmarking purposes, Fig.~\ref{fig:static-perfect} assumes that users have perfect \ac{CSI} measurements at their disposal.
In this deterministic regime, the algorithm converges to a stable \acl{NE} state within a few iterations (for simplicity, we only plotted $4$ users with diverse channel characteristics).
In turn, this rapid convergence leads to drastic gains in energy efficiency, ranging between $3\times$ and $6\times$ over uniform power allocation schemes.

Subsequently, the simulation cycle above was repeated in the presence of observation noise and measurement errors.
The intensity of the measurement noise was quantified via the relative error level of the gradient observations $\hat\payv$, i.e. the standard deviation of $\hat\payv$ divided by its mean (so a relative error level of $z\%$ means that, on average, the observed matrix $\hat\payv$ lies within $z\%$ of its true value).
We then plotted the users' transmit \acl{EE} over time for noise levels $z=25\%$, $50\%$, and $100\%$ (corresponding to moderate, high, and extremely high uncertainty respectively).
Fig.~\ref{fig:static} shows that the network's rate of convergence to a \acl{NE} is negatively impacted by the magnitude of the noise;
remarkably however, \ac{MXL} retains its convergence properties and the network's users achieve a $100\%$ per capita gain in energy efficiency within a few tens of iterations, even under extremely high uncertainty (of the order of $z=100\%$).

%\VB{To be clear: the z\% doesn't mean that at every iteration the estimation is necessarily wrong by z\%. The error is less than or equal to z\%. Am I wrong?}
%\PM{Yes (for the ``am I wrong'' part :P), the above means that the relative error is \emph{on average} $z\%$, so it could be both above and below $z\%$ at any given iteration.}

Finally, to assess the algorithm's performance in a fully dynamic network environment, Fig.~\ref{fig:varying} focuses on mobile users with channels that vary with time due to (Rayleigh) fading, path loss fluctuations, etc.
To simulate this scenario, we used the standard \ac{ETU} model for the users' environment and the \ac{EPA} and \ac{EVA} models to emulate pedestrian ($3$\textendash$5\,\kmh$) and vehicular movement ($30$\textendash$130\,\kmh$) respectively \cite{3GPP}.

In Fig.~\ref{fig:varying-channels}, we plotted the channel gains ($\trof{\bH\bH^{\dag}}$) of $4$ users with diverse mobility and distance characteristics (two pedestrian and two vehicular users, one closer and one farther away from their intended receiver).
As can be seen, the users' channels exhibit significant fluctuations (in the range of a few $\db$) over different time scales, so the \acl{NE} set of the \acl{EE} game described in Section \ref{sec:EE} will evolve itself over time.
Nevertheless, despite the channels' variability, Fig.~\ref{fig:varying-tracking} shows that \ac{MXL} adapts to this highly volatile network environment very quickly, allowing users to track the game's instantaneous equilibrium with remarkable accuracy.
For comparison, we also plotted the users' achieved \acl{EE} under a uniform power allocation policy, which is known to be optimal under isotropic fading conditions \cite{PCL03}.
Because urban environments are not homogeneous and/or isotropic (even on average), uniform power allocation fails to adapt to the changing wireless landscape and performs consistently worse than \ac{MXL} (achieving an \acl{EE} ratio between $2\times$ and $6\times$ lower than that of \ac{MXL}).

%Furthermore, to study the performance of the proposed algorithm in the massive \ac{MIMO} regime, we also plotted in Fig.~\ref{fig:scalability} the number of iterations required for the system to equilibrate for different numbers of antennas at the transmitter and the receiver end ($\tx$ and $\rx$ respectively).
%Importantly, as can be seen in Fig.~\ref{fig:scalability}, the \ac{MXL} algorithm scales very well with the number of antennas and converges within a few iterations, even in the massive \ac{MIMO} regime.
%Specifically, the algorithm requires around $10$ iterations to converge for $8\times92$ \ac{MIMO} systems and around $20$ iterations for $16\times128$ systems.

%----------------------------------------------------------------------
%%% CONCLUSIONS
%----------------------------------------------------------------------
\section{Conclusions and Perspectives}
\label{sec:conclusions}
%----------------------------------------------------------------------
%%% CONCLUSIONS
%----------------------------------------------------------------------
% !TEX root = ./Main.tex

In this paper, we examined a distributed \acl{MXL} algorithm for stochastic semidefinite optimization problems and games that arise in key areas of signal processing and wireless communications (ranging from image-based similarity search to \ac{MIMO} systems and wireless \acl{MAC}).
The main idea of the proposed method is to track the players' individual payoff gradients in a dual, unconstrained space, and then map this process back to the players' action spaces via an ``exponential projection'' step.
Thanks to the aggregation of the players' payoff gradients, the algorithm is capable of operating under uncertainty and feedback noise, two impediments that can have a detrimental effect on more aggressive best-response methods.

To analyze the proposed algorithm, we introduced the notion of a stable \acl{NE}, and we showed that the algorithm is globally convergent to such equilibria \textendash\ or locally convergent when an equilibrium is only locally stable.
Our convergence analysis also revealed that, on average, the algorithm converges to an $\eps$-neighborhood of a \acl{NE} (in terms of the \acl{KL} distance) within $\bigoh(1/\eps)$ iterations.
To validate our theoretical analysis, we also tested the algorithm's performance in realistic multi-carrier/multiple-antenna wireless scenarios where several users seek to maximize their \acl{EE}:
in this setting, users quickly reach a \acl{NE} and attain gains between $100\%$ and $500\%$ in \acl{EE}, even under very high uncertainty.

The above results are particularly promising and suggest that our analysis applies to an even wider setting than the game-theoretic framework \eqref{eq:game} \textendash\ for instance, games with convex action sets that are not necessarily of the form \eqref{eq:feasible}.
Another natural question that arises is whether it is possible to run the proposed \ac{MXL} without \emph{any} gradient information.
%Recent results in simultaneous stochastic approximation \cite{SS11} suggest that this may indeed be possible;
We intend to explore these directions at depth in future work.

%*************************************************************
%*****    APPENDICES
%*************************************************************
\appendices
\numberwithin{equation}{section}
\numberwithin{theorem}{section}
\numberwithin{proposition}{section}
\numberwithin{lemma}{section}

%----------------------------------------------------------------------
%%% ENTROPY
%----------------------------------------------------------------------
\section{The Exponentiation Step}
\label{app:entropy}
%----------------------------------------------------------------------
%%% APPENDIX: ENTROPY
%----------------------------------------------------------------------
% !TEX root = ./Main.tex

In this appendix, our goal will be to establish certain properties of the exponential map of Algorithm \ref{alg:MXL} that are crucial in the stationarity and convergence analysis of the next appendices.
For simplicity, we only treat the case $A_{\play} = 1$;
the general case follows by a trivial rescaling so we do not present it.

With a fair degree of hindsight, we begin by introducing the modified von Neumann entropy \cite{Ved02}:
\begin{flalign}
\label{eq:entropy}
h(\bX)
	&= \trof[\big]{\bX \log\bX}
	+ (1 - \tr(\bX)) \logof{1 - \tr(\bX)}.
\end{flalign}
The convex conjugate of $h$ over the spectrahedron $\spectron = \setdef{\bX\in\psd{M}}{\tr(\bX) \leq 1}$ is then defined as:
\begin{equation}
\label{eq:conj}
h^{\ast}(\bY)
	= \max\setdef{\trof{\bY\bX} - h(\bX)}{\bX\in\spectron},
\end{equation}
with $\bY\in\herm^{\tx}$.
As it turns out, the exponentiation step of the \ac{XL} algorithm is simply the (matrix) derivative of $h^{\ast}$:

\begin{proposition}
\label{prop:choice}
With notation as above, we have:
\begin{equation}
\label{eq:conjugate}
h^{\ast}(\bY)
	= \logof*{1 + \tr(\exp(\bY))},
\end{equation}
and
\begin{equation}
\label{eq:mirror}
\nabla h^{\ast}(\bY)
	= \mirror(\bY)
	\equiv \frac{\exp(\bY)}{1 + \trof{\exp(\bY)}}.
\end{equation}
\end{proposition}

\begin{IEEEproof}
Since the von Neumann entropy is strictly convex \cite{Ved02} and becomes infinitely steep at the boundary of $\feas$, it follows that the maximization problem \eqref{eq:conj} admits a unique solution $\bX\in\spectron^{\circ}$.
Hence, by the first-order \ac{KKT} conditions for the problem \eqref{eq:conj}, we get:
\begin{equation}
\label{eq:KKT}
\bY - \log\bX + \logof{1 - \tr(\bX)} \bI
	= 0.
\end{equation}
%or, after a slight rearrangement:
%\begin{equation}
%\label{eq:KKT-Y}
%e^{\bY}
%	= \frac{\bX}{1 - \tr(\bX)}.
%\end{equation}
Solving for $\bX$ then yields
\begin{equation}
\bX
	= \frac{\exp(\bY)}{1 + \trof{\exp(\bY)}},
\end{equation}
and our claim follows by substituting the above in \eqref{eq:conj}.
\end{IEEEproof}

%\begin{remark}
%For the case $A_{\play} = \infty$, we need instead to consider the entropy function $h(\bX) = \trof{\bX \log\bX - \bX}$.
%In turn, this gives $h^{\ast}(\bY) = \trof{\exp(\bY)}$ and $\mirror(\bY) = \exp(\bY)$ instead of the expressions \eqref{eq:conjugate} and \eqref{eq:mirror} respectively.
%\end{remark}

In addition to the above, the von Neumann entropy also provides a ``congruence'' measure between the primal variables $\bX$ and the auxiliary ``dual'' variables $\bY$.
Specifically, following \cite{MS16}, we introduce here the \emph{Fenchel coupling}:
\begin{equation}
\label{eq:Fenchel}
\fench(\bX,\bY)
	= h(\bX) + h^{\ast}(\bY) - \trof{\bY\bX}
	= \breg(\bX,\mirror(\bY)).
\end{equation}
By Fenchel's inequality, we have $\fench(\bX,\bY) \geq 0$ with equality if and only if $\bY = \nabla h(\bX)$ \textendash\ or, equivalently, iff $\bX = \mirror(\bY)$.
More importantly for our purposes, we also have the following approximation lemma:

\begin{proposition}
\label{prop:Fenchel-approx}
For all $\bX\in\spectron$ and for all $\bY,\bZ\in\herm^{\tx}$, we have
\begin{equation}
\label{eq:Fenchel-approx}
\fench(\bX,\bY + \bZ)
	\leq \fench(\bX,\bY)
	+ \trof{\bZ\,(\mirror(\bY) - \bX)}
	+ \dnorm{\bZ}^{2}.
\end{equation}
\end{proposition}

\begin{IEEEproof}
By the definition of the Fenchel coupling, we get:
\begin{flalign}
\fench(\bX,\bY+\bZ)
	&= h(\bX) + h^{\ast}(\bY + \bZ) - \trof{(\bY + \bZ) \bX}
	\notag\\
	&\leq h(\bX) + h^{\ast}(\bY) + \trof{\bZ\,\mirror(\bY)} + \dnorm{\bZ}^{2}
	\notag\\
	&- \trof{\bY\bX} - \trof{\bZ\bX}
	\notag\\
	&= \fench(\bX,\bY) + \trof{\bZ (\mirror(\bY) - \bX)} + \dnorm{\bZ}^{2},
\end{flalign}
where the expansion of $h^{\ast}$ in the second line follows from the fact that the von Neumann entropy is $1/2$-strongly convex (from the duality of strong convexity and strong smoothness, and the fact that $h^{\ast}$ is $2$-strongly smooth) \cite{KSST12}.
\end{IEEEproof}

%----------------------------------------------------------------------
%%% STATIONARITY
%----------------------------------------------------------------------
\section{Stationarity Analysis}
\label{app:stationary}
%----------------------------------------------------------------------
%%% APPENDIX: STATIONARITY
%----------------------------------------------------------------------
% !TEX root = ./Main.tex
We prove Theorem \ref{thm:stationary} regarding the possible termination states of Algorithm \ref{alg:MXL}:

\begin{IEEEproof}[Proof of Theorem \ref{thm:stationary}]
Let $\payveq = \payv(\eq)$ and assume that $\eq$ is not a Nash equilibrium.
By Eq.~\eqref{eq:Nash}, this implies that $\trof{(\bX_{\play}' - \eq_{\play}) \payveq_{\play}} > 0$ for some player $\play\in\playset$ and some $\bX'\in\feas_{\play}$.
Therefore, by continuity, there exists some $a>0$ such that
\begin{equation}
\label{eq:notNash}
\trof{(\bX_{\play}' - \bX_{\play})\,\payv_{\play}''}
	\geq a > 0,
\end{equation}
for all $\bX$ in a small enough neighborhood $\nhd$ of $\eq$ in $\feas$ and for all $\payv_{\play}''$ sufficiently close to $\payveq_{\play}$.

Since $\bX(n) \to \eq$ as $n\to\infty$, we may assume that $\bX(n) \in \nhd$ for all $n$ (note that the step-size condition $\sum_{n=1}^{\infty} \step_{n}^{2} < \sum_{n=1}^{\infty} \step_{n} = \infty$ is not affected if we start the sequence at some finite $n_{0}>0$).
The recursion \eqref{eq:MXL} then yields:
\begin{equation}
\label{eq:stat1}
\bY(n)
	= \bY(0) + t_{n} \bar\payv(n),
\end{equation}
where we have set $t_{n} = \sum_{j=1}^{n} \step_{j}$ and
\begin{equation}
\label{eq:stat2}
\bar\payv(n)
	= \frac{1}{t_{n}} \sum_{j=1}^{n} \step_{j} \hat\payv(j)
	= \frac{1}{t_{n}} \sum_{j=1}^{n} \step_{j} \payv(\bX(j))
	+ \frac{1}{t_{n}} \sum_{j=1}^{n} \step_{j} \noise(j)
\end{equation}
denotes the $\step$-weighted time average of the received gradient estimates $\hat\payv(n)$.
By the strong law of large numbers for martingale differences \cite[Theorem 2.18]{HH80}, we get $\lim_{n\to\infty} n^{-1} \sum_{j=1}^{n} \noise(j) = 0$ (a.s.), so the last term of \eqref{eq:stat2} also converges to zero (a.s.) by Hardy's summability criterion \cite[Theorem 14]{Har49} applied to the weight sequence $w_{j,n} = \step_{j}/t_{n}$.
Thus, given that $\bX(n) \in \nhd$ for all $n$, we conclude that $\bar\payv(n)\to\payveq$ as $n\to\infty$.

Now, with notation as in Appendix \ref{app:entropy}, let $h_{\play}(\bX_{\play}) = \trof{\bX_{\play} \log\bX_{\play}} + (1 - \tr(\bX_{\play})) \log(1 - \tr(\bX_{\play}))$.
Since $\bX_{\play}(n) = \nabla h_{\play}^{\ast}(\bY_{\play}(n))$ by Proposition \ref{prop:choice}, we will also have $\nabla h_{\play}(\bX_{\play}(n)) = \bY_{\play}(0) + t_{n} \bar\payv_{\play}(n)$ by the general theory of convex conjugation.
In turn, this implies that
\begin{equation}
\label{eq:stat3}
h_{\play}(\bX_{\play}') - h_{\play}(\bX_{\play}(n))
	\geq \trof{(\bY_{\play}(0) + t_{n}\bar\payv_{\play}(n))\,(\bX_{\play}' - \bX_{\play}(n))},
\end{equation}
by the convexity of $h_{\play}$.
However, since $\lim_{n\to\infty} \bar\payv_{k}(n) = \payveq_{\play}$ and $\lim_{n\to\infty} t_{n} = \infty$, Eq.~\eqref{eq:notNash} yields
\begin{equation}
\label{eq:stat3}
h_{\play}(\bX_{\play}') - h_{\play}(\bX_{\play}(n))
	\gtrsim a t_{n},
\end{equation}
so $h_{\play}(\bX_{\play}') - h_{\play}(\bX_{\play}(n)) \to \infty$ as $n\to\infty$, a contradiction.
\end{IEEEproof}

%----------------------------------------------------------------------
%%% GLOBAL
%----------------------------------------------------------------------
\section{Global Convergence}
\label{app:global}
%----------------------------------------------------------------------
%%% APPENDIX: GLOBAL
%----------------------------------------------------------------------
% !TEX root = ./Main.tex

We begin our analysis with an auxiliary result for the convergence of \eqref{eq:MXL} in continuous time.
Specifically, consider the dynamics
\begin{equation}
\label{eq:MXL-cont}
\tag{\ref*{eq:MXL}$_{c}$}
\begin{aligned}
\dot\bY
	&= \payv(\bX),
	\\
\bX
	&= \mirror(\bY),
\end{aligned}
\end{equation}
obtained by taking the continuous-time limit of \eqref{eq:MXL}.
Our first auxiliary result is that globally stable states are globally attracting under \eqref{eq:MXL-cont}:

\begin{proposition}
Let $\eq$ be a globally stable \acl{NE} and let $\bX(t)$ be a solution of \eqref{eq:MXL-cont}.
Then, $\lim_{t\to\infty} \bX(t) = \eq$.
\end{proposition}

\begin{IEEEproof}
Let $\lyap(t) = \fench(\eq,\bY(t))$.
A simple differentiation then yields
\begin{equation}
\label{eq:Lyapunov}
\dot \lyap
	= \trof{\dot\bY \, \nabla h^{\ast}(\bY)} - \trof{\dot\bY \, \eq}
	= \trof{(\bX - \eq) \, \payv(\bX)},
\end{equation}
i.e. $\dot H \leq 0$ with equality if and only if $\bX = \eq$ (recall that $\eq$ is assumed globally stable).
This implies that $H(t)$ is nonincreasing, and hence converges to some $c\geq0$ as $t\to\infty$.
%Now, since the sublevel sets $\sublvl_{\eps}(\eq) = \setdef{\bX=\mirror(\bY)\in\feas}{\fench(\eq,\bY) \leq \eps}$ are bounded,
%%\footnote{It is easy to check that this is true even if $A_{\play}=\infty$, i.e. if $\feas$ is not compact.}
%it follows that $\bX(t)$ is contained in a compact subset of $\feas$.
Hence, by compactness, there exists some $\hat\bX\in\feas$ and a sequence $t_{n}\nearrow\infty$ such that $\bX(t_{n}) \to \hat\bX$ as $n\to\infty$.

Assume now {ad absurdum} that $\hat\bX\neq\eq$, so there exists some $a>0$ and a neighborhood $\nhd$ of $\hat\bX$ such that $\trof{(\bX - \eq)\,\payv(\eq)} \leq -a$ for all $\bX\in\nhd$;
furthermore, since $\norm{\dot\bX}$ is bounded from above (recall that $\mirror$ is Lipschitz), there exists some $\delta>0$ such that $\bX(t) \in U$ for all $t\in[t_{n},t_{n}+\delta]$ and for all $n\geq0$.
In that case however, \eqref{eq:Lyapunov} yields:
\begin{flalign}
\lim_{t\to\infty} H(t)
	&\leq H(0) + \sum_{n=1}^{\infty} \int_{t_{n}}^{t_{n}+\delta} \trof{(\bX(t) - \eq)\,\payv(\bX(t))} \dd t
	\notag\\
	&\leq H(0) - \sum_{n=1}^{\infty} a\delta
	= - \infty,
\end{flalign}
a contradiction.
This shows that $\eq$ is the only potential $\omega$-limit point of $\bX(t)$;
since $\bX(t)$ admits at least one $\omega$-limit, it follows that $\bX(t)\to\eq$, as claimed.
\end{IEEEproof}

%We now proceed to show that the iterates of \eqref{eq:MXL} are asymptotically close to solution segments of \eqref{eq:MXL-cont} of arbitrary length \textendash\ more precisely, that they comprise an \emph{\acl{APT}} of \eqref{eq:MXL-cont} in the sense of \cite{Ben99}.
%
%\begin{proposition}
%Assume that \eqref{eq:MXL} is run with a nonincreasing step-size sequence $\step_{n}$ such that $\sum_{n} \step_{n}^{2} < \sum_{n} \step_{n} = +\infty$ and noisy measurements $\hat\payv_{k}$ satisfying \eqref{eq:zeromean} and \eqref{eq:MSE}.
%Then, the iterates $\bX(n)$ of \eqref{eq:MXL} form an \acl{APT} of \eqref{eq:MXL-cont}.
%\end{proposition}
%
%\begin{IEEEproof}
%Simply note that the recursion \eqref{eq:MXL} can be written in the form:
%\begin{equation}
%\label{eq:Y0}
%\bY(n+1)
%	= \bY(n)
%	+ \step_{n} \left[ \payv(\bX(n)) + \bZ(n) \right].
%\end{equation}
%Since the map $\bY\mapsto\bX$ is Lipschitz continuous%
%\footnote{This follows from the fact that the von Neumann entropy \eqref{eq:entropy} is strongly convex with respect to the nuclear norm \cite{KSST12,Nes09}.}
%and the rate function $\rate(\bX)$ is smooth over $\spectron$, it follows that the map $\bY\mapsto\payv(\bX(\bY))$ is Lipschitz and bounded.
%Our claim then follows from Propositions 4.2 and 4.1 in \cite{Ben99}.
%\end{IEEEproof}

With this auxiliary result at hand, we are finally in a position to prove our global convergence result:

\begin{IEEEproof}[Proof of Theorem \ref{thm:global}]
We first note that the recursion \eqref{eq:MXL} can be written in the more succinct form:
\begin{equation}
\label{eq:Y0}
\bY(n+1)
	= \bY(n)
	+ \step_{n} \left[ \payv(\mirror(\bY(n))) + \bZ(n) \right].
\end{equation}
Since $\payv(\bX)$ is differentiable for almost all $\bX\in\feas$ by Alexandrov's theorem, Propositions 4.1 and 4.2 in \cite{Ben99} show that $X(n)$ is an \emph{asymptotic pseudotrajectory} of \eqref{eq:MXL-cont}, i.e. the iterates of \eqref{eq:MXL} are asymptotically close to solution segments of \eqref{eq:MXL-cont} of arbitrary length \textendash\ for a precise statement, see \cite[Sec.~3]{Ben99}.

Assume now that $\bX(n)$ remains at a minimal positive distance from $\eq$.
Since $\eq$ is globally stable, we will have $\trof{(\bX(n) - \eq) \, \payv(\bX(n))} \leq -a$ for some $a>0$ and for all $n$.
Furthermore, if we let $D_{n} = \fench(\bX,\bY(n))$, Proposition \ref{prop:Fenchel-approx} yields:
\begin{flalign}
\label{eq:Dn0}
D_{n+1}
	&
%	= \fench(\eq,\bY(n+1))
	= \fench(\eq,\bY(n) + \step_{n} \hat\payv(n))
%	&\leq \fench(\eq,\bY(n))
%	+ \step_{n} \tr\left[ \hat\payv(n) \cdot \nabla_{\bY(n)} \fench(\eq,\bY(n)) \right]
%	+ \frac{1}{2} \step_{n}^{2} \dnorm{\hat\payv(n)}^{2}
	\notag\\
	&\leq D_{n}
%	+ \step_{n} \tr[(\bX(n) - \eq) \, \payv(\bX(n))]
	+ \step_{n} v_{n}
	+ \step_{n} \xi_{n}
	+ \step_{n}^{2} \dnorm{\hat\payv(n)}^{2},
\end{flalign}
where we have set $v_{n} = \tr[(\bX(n) - \eq) \, \payv(\bX(n))]$ and $\xi_{n} = \tr[ (\bX(n) - \eq) \, \bZ(n)]$.
Hence, telescoping \eqref{eq:Dn0} yields:
\begin{equation}
\label{eq:Dn1}
\txs
D_{n+1}
	\leq D_{0}
	- t_{n} \left( a - \sum_{j=1}^{n} w_{j,n}\,\xi_{j} \right)
	+ \sum_{j=1}^{n} \step_{j}^{2} \dnorm{\hat \payv(j)}^{2},
\end{equation}
where $t_{n} = \sum_{j=1}^{n} \step_{j}$ and $w_{j,n} = \step_{j}/t_{n}$.
By the strong law of large numbers for martingale differences \cite[Theorem 2.18]{HH80}, we have $n^{-1} \sum_{j=1}^{n} \xi_{j} \to0$ (a.s.);
hence, given that $\step_{n+1}/\step_{n}\leq1$, Hardy's summability criterion \cite[Thm.~14]{Har49} applied to the sequence $w_{j,n} = \step_{j}/t_{n}$ yields $\sum_{j=1}^{n} w_{j,n}\, \xi_{j} \to 0$ \as.
Finally, since $\step_{n}$ is square-summable and $\step_{n}\bZ(n)$ is a martingale difference with finite variance, Theorem 6 in \cite{Cho68} shows that $\sum_{n=1}^{\infty} \step_{n}^{2} \dnorm{\hat\payv(n)}^{2} < \infty$ \as.

Since $t_{n}\to\infty$ by assumption, the above implies that the RHS of \eqref{eq:Dn1} tends to $-\infty$ \as;
this contradicts the fact that $D_{n}\geq0$, so we conclude that $\bX(n)$ visits a compact neighborhood of $\eq$ infinitely often (viz. there exists a sequence $n_{k}\nearrow\infty$ such that $\bX(n_{k})$ lies in said neighborhood).
Since $\eq$ attracts any initial condition $\bX(0) = \mirror(\bY(0))$ under the continuous-time dynamics \eqref{eq:MXL-cont}, Theorem 6.10 in \cite{Ben99} then shows that $\bX(n)$ converges to $\eq$ \as, as claimed.
\end{IEEEproof}

%----------------------------------------------------------------------
%%% LOCAL
%----------------------------------------------------------------------
\section{Local Convergence}
\label{app:local}
%----------------------------------------------------------------------
%%% APPENDIX: LOCAL
%----------------------------------------------------------------------
% !TEX root = ./Main.tex

\begin{IEEEproof}[Proof of Theorem \ref{thm:local}]
By the definition of local stability, there exists a neighborhood $U$ of $\eq$ in $\feas$ such that \eqref{eq:VS} holds for all $\bX\in U$.
Assume now that $m>0$ is taken sufficiently small so that $\mirror(\bY) \in U$ whenever $\fench(\eq,\bY) < 4m$ (the existence of such a positive $m$ follows from the fact that $\mirror(\bY)\to\eq$ if $\fench(\eq,\bY)\to0$).
By Eq.~\eqref{eq:Lyapunov}, it then follows that the set $U_{4m} = \setdef{\bY}{\fench(\eq,\bY) \leq 4m}$ is invariant under \eqref{eq:MXL-cont}.
Hence, by shadowing the proof of Theorem \ref{thm:global}, it suffices to show that there exists an open set $U_{\eps} \subseteq U_{4m}$ such that $\probof{\text{$\bY(n) \in U_{4m}$ for all $n$}} \geq 1-\eps$ whenever $\bY(0)\in U_{\eps}$.

To that end, let $D_{n} = \fench(\eq,\bY(n))$ and assume that $\fench(\eq,\bY(0)) \leq m$.
Then, \eqref{eq:Dn0} yields
\begin{flalign}
\label{eq:Dn-local}
\!
D_{n}
	&\leq m
	+ \sum_{j=1}^{n} \step_{j} v_{j}
	+ \sum_{j=1}^{n} \step_{j} \xi_{j}
	+ \sum_{j=1}^{n} \step_{j}^{2} \dnorm{\hat\payv(j)}^{2}
	\notag\\
	&\leq m
	+ \sum_{j=1}^{n} \step_{j} v_{j}
	+ \sum_{j=1}^{n} \step_{j} \xi_{j}
	+ \sum_{j=1}^{n} \step_{j}^{2} \left( L^{2} + K \dnorm{\bZ(j)}^{2} \right),
\end{flalign}
where $v_{j} = \trof{(\bX(j) - \eq)\,\payv(j)}$, $\xi_{j} = \trof{(\bX(j) - \eq)\,\bZ(j)}$, $K$ is a sufficiently large positive constant, and $L = \sup_{\bX\in\feas} \dnorm{\payv(\bX)}$.
By the subexponential moment growth condition \eqref{eq:subexp} and the martingale concentration inequalities of \cite[Theorem 1.2A]{dlP99}, it follows that $\probof{\sum_{j=1}^{n} \step_{j} \xi_{j} \geq m} \leq \exp\left(-\frac{m^{2}}{2\noisevar(\sum_{j=1}^{n}\step_{j}^{2} + \step_{1} m/\noisedev)}\right) \leq \eps$ if $\step$ is chosen sufficiently small (specifically, so that $\step_{1} m/\noisedev + \sum_{j=1}^{\infty} \step_{j}^{2} \leq m^{2} \abs{\log \eps}^{-1}$).
Likewise, \eqref{eq:subexp} implies that the tails of $\bZ$ are subexponential, so there exists some constant $A>0$ such that $\probof{\dnorm{\bZ(j)} \geq m/\step_{j}} \leq A \exp(- m/\step_{j})$.
By the summability assumption for $\step_{j}$ and the Borel-Cantelli lemma, it follows that $\probof[\big]{\dnorm{\bZ(j)} \geq m/\step_{j}\;\text{for infinitely many $j$}} = 0$, so the last term of \eqref{eq:Dn-local} is bounded from above by $m$ \as if $\step$ is chosen small enough.
Finally, by choosing $\step$ sufficiently small so that $\sum_{j=1}^{\infty} \step_{j}^{2} \leq m/L^{2}$, the third term of \eqref{eq:Dn-local} is also bounded by $m$.

Combining all of the above, we obtain
%\begin{equation}
\(
D_{n}
	\leq 4m + \sum_{j=1}^{n} \step_{j} v_{j}
\)
%\end{equation}
with probability at least $1-\eps$.
Since $\mirror(U_{4m}) \subseteq U$ by construction, we have $v_{j} \leq 0$ for all $j=1,\dotsc,n$, and we conclude that $D_{n} \leq 4m$ with probability at least $1-\eps$.
This implies that $\probof{\bY(n) \in U_{4m} \; \text{for all $n$}} \leq 1-\eps$, as claimed.
\end{IEEEproof}

%----------------------------------------------------------------------
%%% RATES
%----------------------------------------------------------------------
\section{Rates of Convergence}
\label{app:rates}
%----------------------------------------------------------------------
%%% APPENDIX: CONVERGENCE RATE
%----------------------------------------------------------------------
% !TEX root = ./Main.tex

In this last appendix, our goal is to derive the convergence rate of \acl{MXL}:

\begin{IEEEproof}[Proof of Theorem \ref{thm:convrate}]
Let $\bar D_{n} = \exof{\fench(\eq,\bY(n))}$.
Then, taking expectations in \eqref{eq:Dn0} yields
\begin{flalign}
\label{eq:Dn-mean}
\bar D_{n+1}
	&\leq \bar D_{n}
	+ \step_{n} \exof{\trof{(\bX(n) - \eq) \, \payv(n)}}
	+ \step_{n}^{2} \exof{\dnorm{\hat\payv(n)}^{2}}
	\notag\\
	& \leq (1 - \step_{n} B) \bar D_{n}
	+ \step_{n}^{2} \vbound^{2},
\end{flalign}
where, in the second line, we used \eqref{eq:vbound} and the assumption that $\eq$ is $B$-strongly stable.
%and the second moment bound \eqref{eq:vbound}.
With this in mind, assume inductively that $\bar D_{n} \leq A/n$ for some $A > 0$;
it will then suffice to show that $\bar D_{n+1} \leq A/(n+1)$ and that we can choose $A  = \step^{2} \vbound^{2}/(B\step - 1)$.
Therefore, substituting in \eqref{eq:Dn-mean}, it suffices to show that
\begin{equation}
\left( 1 - \frac{B\step}{n}\right) \frac{A}{n} + \frac{\step^{2} \vbound^{2}}{n^{2}}
	\leq \frac{A}{n+1}.
\end{equation}
Rearranging this last equation, we get
\begin{equation}
nA
	\leq (n+1) (AB\step - \step^{2} \vbound^{2}),
\end{equation}
which shows that it suffices to pick $A = AB\step - \step^{2} \vbound^{2}$.
Solving for $A$ then yields $A = \step^{2} \vbound^{2} / (B\step - 1)$, as claimed.

Finally, the bound \eqref{eq:convrate-ext} follows by noting that $\breg(\eq,\bX) =\Omega(\norm{\eq - \bX})$ if $\eq$ is an extreme point of $\feas$;
likewise, the bound \eqref{eq:convrate-int} follows by noting that $\breg(\eq,\bX) =\Omega(\norm{\eq - \bX}^{2})$ if $\eq$ is interior, and subsequently applying Jensen's inequality.
\end{IEEEproof}

%*************************************************************
%*****    BIBLIOGRAPHY
%*************************************************************
%\balance
\bibliographystyle{IEEEtran}
\bibliography{IEEEabrv,Bibliography-XL,Bibliography-Examples}

\end{document}